\newcommand{\xg}{$x_\gamma$}
\documentclass{ws-p8-50x6-pb}
\def\JP{{ J. Phys.\ }}

\def\EPJ{ Eur.\ Phys.\ J.\ }

\def\NP{{ Nucl.\ Phys.\ }}

\def\PR{{ Phys.\ Rev.\ }}
\def\ZP{{ Z. Phys.\ }}
\newcommand{\moreht}{\rule[-0ex]{0ex}{4.5ex}}
\newcommand{\nsubj}{$<\!\!n_{subjet}\!\!>$}
\newcommand{\ET}{$E_T$} 
\begin{document}

\title{Photoproduction at low $Q^2$}

\author{P. J. Bussey}

\address{Department of Physics and Astronomy,\\ University of Glasgow,\\
Glasgow G12 8QQ, U.K.\\E-mail: p.bussey@physics.gla.ac.uk\\[4mm]
\mbox{\rm Talk given at 19{\it{th} Physics in Collision\/} conference, Ann Arbor, USA, June 1999.}
}

\maketitle

\enlargethispage{10mm}

\abstracts{
The past year has seen a number of important developments in 
hard photoproduction physics at the HERA collider.     
These are surveyed.
}
\vspace*{-75mm}\hspace{-20mm}
\mbox{{\bf GLAS-PRE/1999-15}\hspace*{9cm}\raisebox{-25mm}{
\epsfig{file=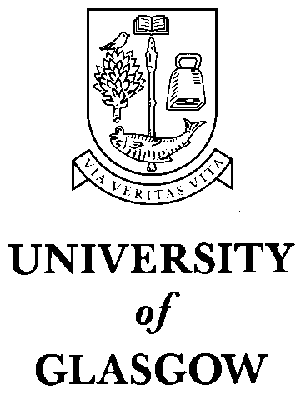,height=31mm}}}\\[35mm]

\section{Introduction}
The record luminosities obtained in recent years at the HERA collider 
have given many new results
in deep inelastic scattering.  However most of the virtual
photons that mediate the $ep$ collisions have low 
$Q^2$ values; these large fluxes of  quasi-real photons have made 
HERA an outstanding laboratory in which to
study photon physics.  Here I survey some of the recent
results reported by the experiments H1 and ZEUS, concentrating
on the hard interactions which reveal the partonic structure of the 
photon.

Throughout, it will be helpful to have in mind three characteristic
modes by which a photon can interact with a proton at high energies.
These are
illustrated in fig.\ 1.\cite{anom}  The simplest is when the entire
photon couples in a pointlike manner to a high transverse energy
($E_T$) $q\bar q$ pair; such diagrams are referred to as ``direct''
photon processes.  The photon may also couple
non-perturbatively to a hadronic state which then interacts with the
proton.  Both the hadronic state and the proton can be sources
of partons which undergo hard QCD scattering; such diagrams are known
as ``resolved'' photon processes.  

These two classes of process are fully separable only in lowest order
(LO) QCD.  A more complex situation occurs when the photon
couples in a pointlike way to a medium-$E_T$ $q\bar q$ pair, one of
which then undergoes a hard scatter.  Here there are three
perturbative couplings and the process is no longer LO.  Such
processes are referred to as ``anomalous''.  Both pointlike and
hadronic photon couplings are also present, of course, in higher order
diagrams.

\begin{figure}[t]\mbox{
\epsfig{file=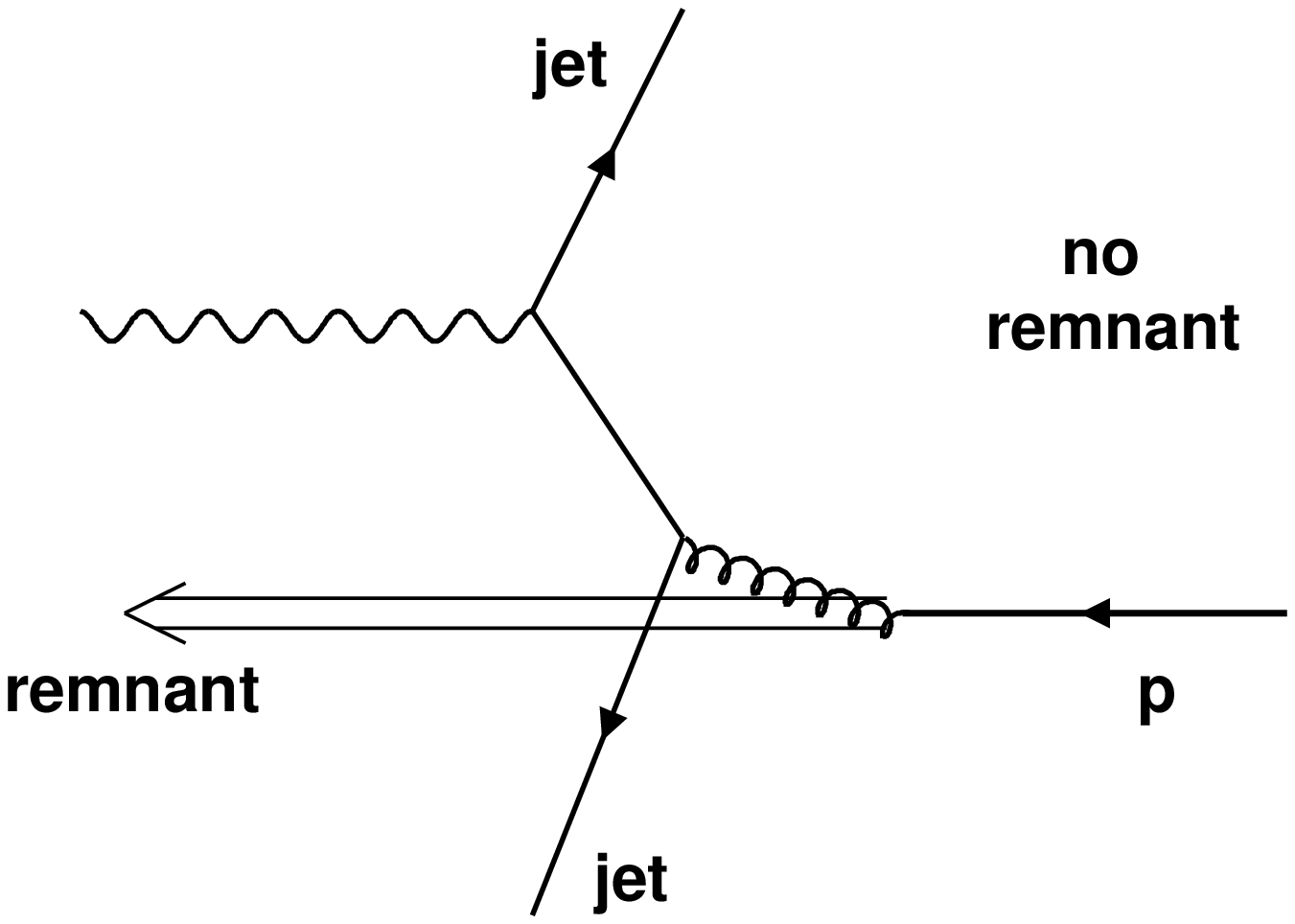,width=4.5cm,%
bbllx=120pt,bblly=90pt,bburx=520pt,bbury=380pt,clip=yes}\hspace*{10mm}
\parbox[b]{5.5cm}{\small (a) Pointlike photon coupling to a high-$E_T$ 
quark pair.  There is no photon remnant. In the LO diagram illustrated, 
\xg\ = 1.\\[2ex]}}\\[1mm]
\mbox{
\epsfig{file=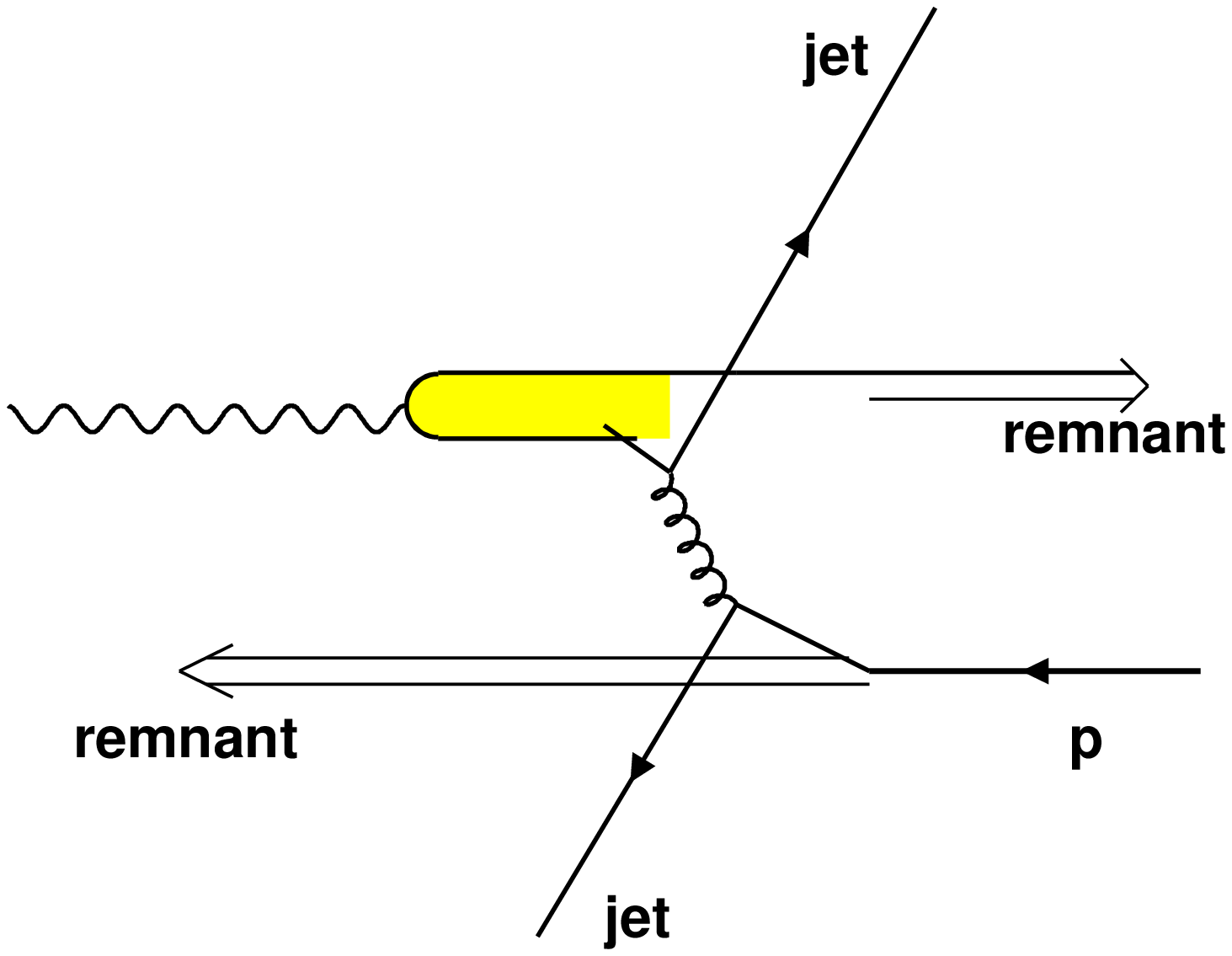,width=4.5cm,%
bbllx=100pt,bblly=90pt,bburx=520pt,bbury=420pt,clip=yes}\hspace*{10mm}
\parbox[b]{5.5cm}{\small (b) Hadronic photon coupling.  The photon acts
as a source of partons which can scatter off partons in the proton.
There is a photon remnant as well as a proton remnant.\\[2ex]}
}\\[1mm]\mbox{
\epsfig{file=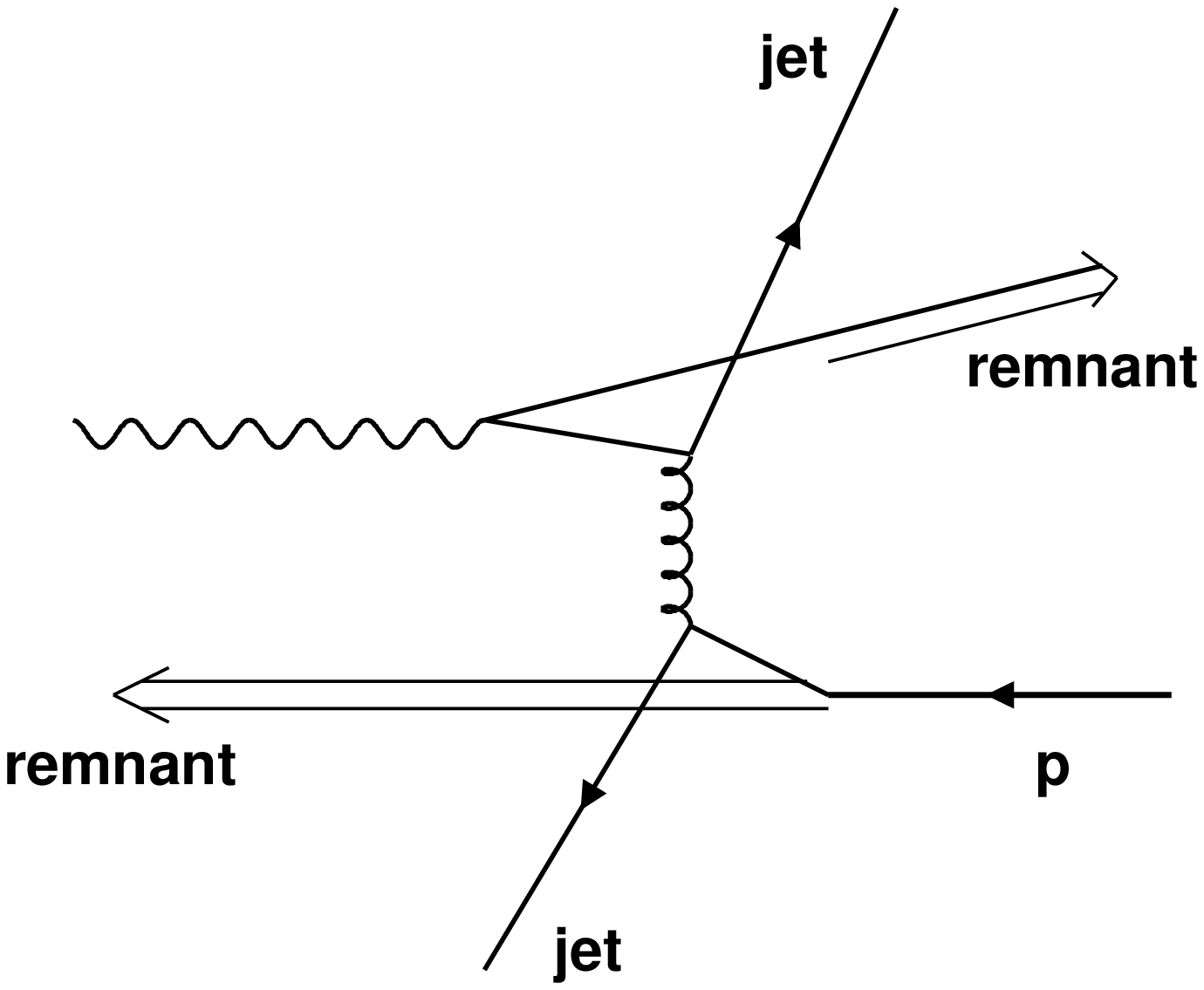,width=4.5cm,%
bbllx=120pt,bblly=90pt,bburx=520pt,bbury=430pt,clip=yes}\hspace*{10mm}
\parbox[b]{5.5cm}{\small (c) Anomalous photon coupling.
The photon couples perturbatively to a medium-$E_T$ quark pair.
There is a photon remnant at non-zero transverse momentum.\\[2ex]}}
\caption{Schematic illustrations of photoproduction processes.}\end{figure}

With the above theoretical ideas in the background, in practice we
wish to make experimental observations.  The principal measurable
quantities available in hard photon reactions are high \ET\ jets, high
\ET\ photons (``prompt'' photons), various single
particle spectra and the remnants of the incoming photon and proton.
The quantities we would like to study include the nature of the
outgoing quarks and gluons, and the hadronic structure of the photon,
in addition to the challenge of verifying that we have an adequate
understanding of the basic QCD mechanisms that are operating.

A basic parton-level concept is that of \xg, the fraction of the
photon energy that takes part in the main hard QCD subprocess.  (Above
LO this may not always be defined simply.)  In practice, what is
wanted is a measurable quantity which correlates with \xg\ in a chosen
theoretical description of the process.  For dijet final states, two
such experimental estimators are
$$ x_\gamma^{\;obs} = 
\frac{\Sigma_{jets}E_T^{\;jet} e^{-\eta^{jet}}}{2 E_\gamma}
\mbox{\hspace*{7mm}and\hspace*{7mm}}  
x_\gamma^{jets} = \frac{\Sigma_{jets}(E^{jet}-p_z^{jet})}
{\Sigma_{hadr}(E^{hadr}-p_z^{hadr})}$$ 
which have been used by ZEUS and H1.  Here as elsewhere,
$\eta$ denotes laboratory pseudorapidity and the proton beam defines the
``forward'' ($+Z$) direction. ($E_T e^{-\eta}\equiv(E - p_z)$; 
the above formulae differ in the averaging of the parameters over the
particles in the jet.)  To facilitate comparisons with theory, both
estimators are defined at the final-state hadron level.

In the following sections, photoproduction
results will be presented on:\\ (i) dijet final states\\
(ii) prompt photons\\(iii) the parton content of the photon\\
(iv) the photon remnant, and\\ (v) properties of the final-state jet system.

\begin{figure}[b!]\centerline{
\raisebox{-3mm}{\epsfig{file=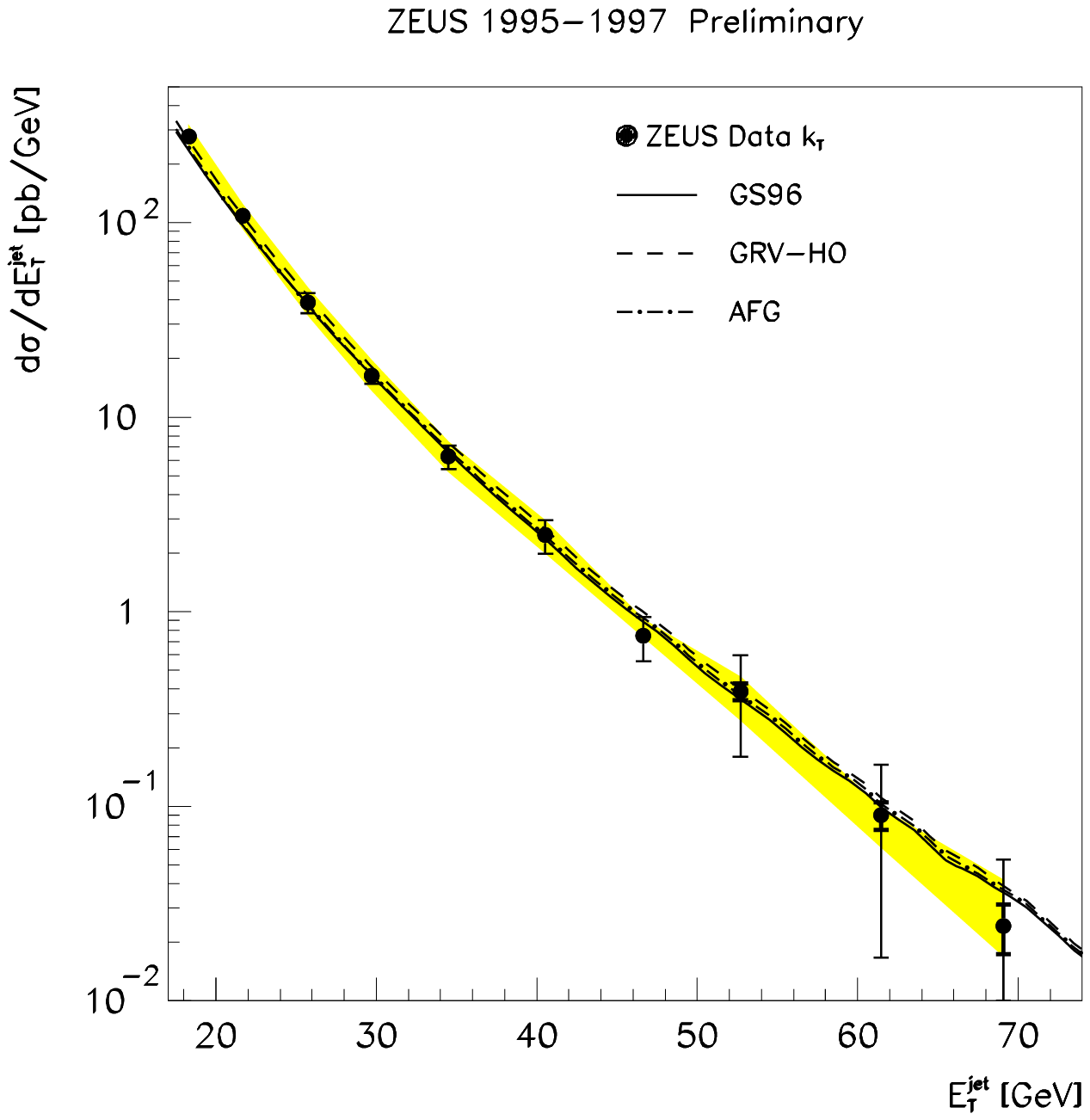,width=5.9cm%
,bbllx=110pt,bblly=310pt,bburx=475pt,bbury=690pt,clip=yes}} 
\hspace*{2mm}\epsfig{file=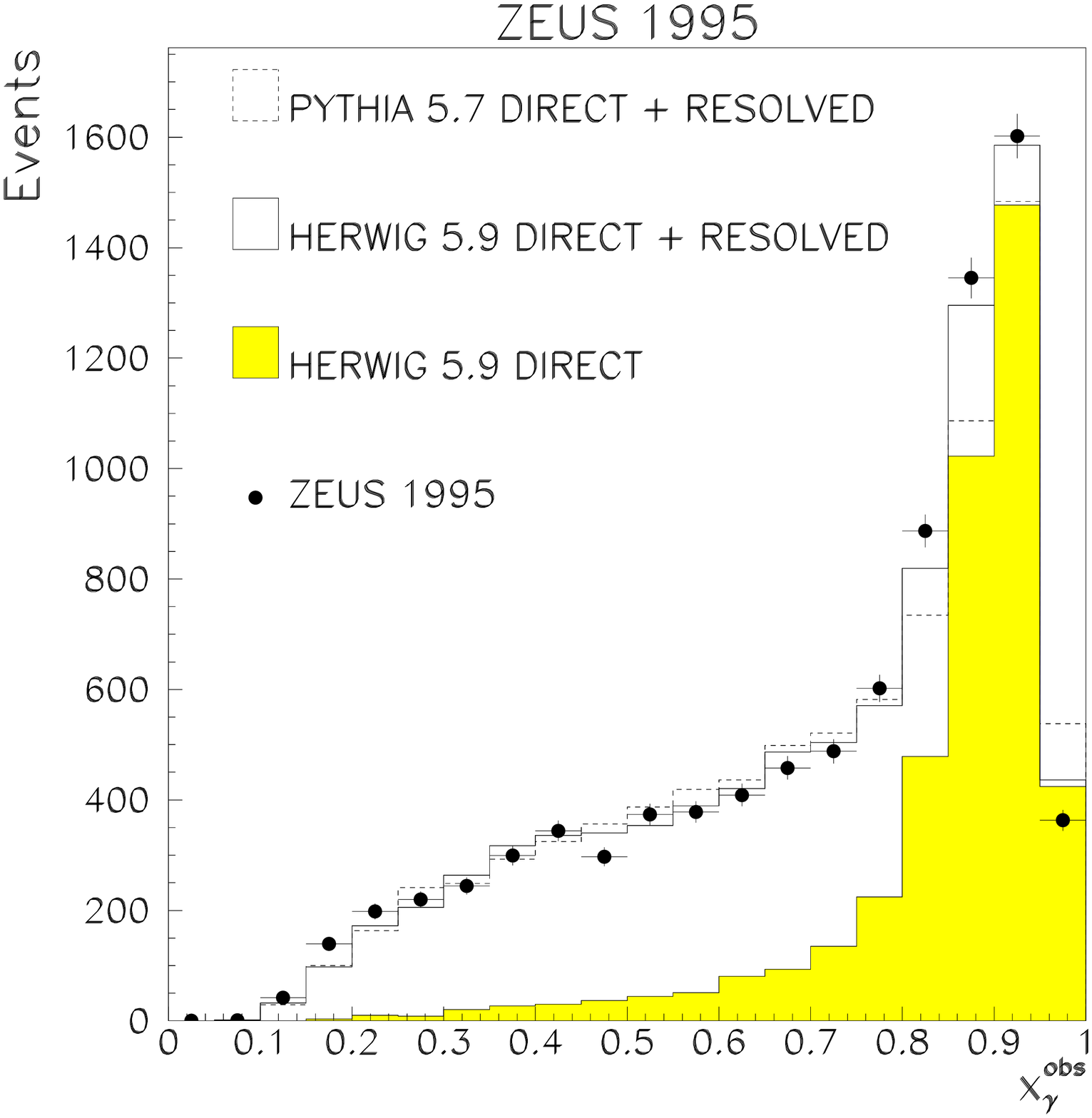,width=5.75cm}\hspace*{1mm} 
}
\vspace*{-4mm}\hspace*{3.1cm}(a)\hspace*{5.9cm}(b)
\caption{(a) Distribution of transverse energy of photoproduced jets in 
ZEUS in the pseudorapidity range $-0.75 < \eta < 2.5$.  (b) Distribution of
$x_\gamma^{obs}$ in dijet events.}\end{figure}

\section{Dijet final states.} 
The integrated luminosities now available from HERA allow jets to be
measured out to high $E_T$ values.  This is illustrated in fig.\ 2(a),
where the \ET\ spectrum seen in ZEUS\cite{yuji} is compared to theory
using several models of the photon parton density, making use of
recent next-to-leading order (NLO) calculations~\cite{ptheory} whose
predictions do not differ from each other on the scale of the
figure. Agreement is excellent in all cases.  The $k_T$ jet algorithm
is used,\cite{KT} in order to avoid the uncertainties concerning seed
definition and jet merging present with the cone algorithm
approach.\cite{ZEUSJETS} Figure 2(b) shows the $x_\gamma^{obs}$
spectrum obtained by ZEUS using high $E_T$ jet pairs.  The shape of
the distribution can be well fitted using PYTHIA and HERWIG LO
simulations of the resolved and direct processes, provided that each
contribution is rescaled.  It is important to note that the shape of
the distribution depends strongly on the cuts imposed upon the jet
parameters, as will be seen below.

By selecting on $x_\gamma^{obs} < 0.75 $ or $x_\gamma^{obs}> 0.75 $,
event samples can be obtained in which the hadronic (resolved) or
pointlike (direct) photon diagrams dominate.

\begin{figure}[t!]
\centerline{\epsfig{file=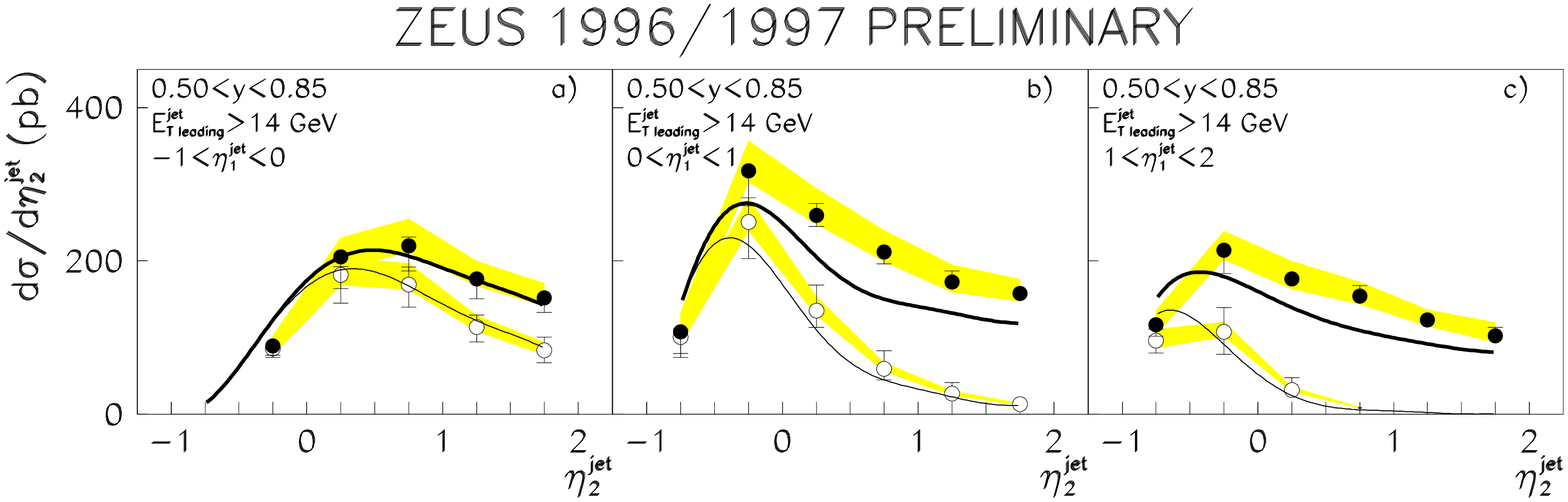,width=12.1cm}}
\caption{Cross sections for photoproduced jets, in dijet events with
minimum jet $E_T$ values of 14 GeV.  The incident photon energy, as a
fraction $y$ of the $e$ beam energy, is restricted to high values.  Full circles =
all $x_\gamma^{obs}$; open circles = $x_\gamma^{obs}>$ 0.75.  
The pseudorapidity of one jet
is plotted for given intervals of that of the other.  The shaded bands
represent experimental uncertainty due to the simulation of the response 
of the ZEUS calorimeter. Uncertainties of up to 15\% due to the
QCD scale, and 10\% due to effects of hadronisation, should be allowed
on the theoretical predictions. }\end{figure}

   The availability of good jet statistics at high $E_T$ removes most
of the problems concerning a possible ``underlying event'', due to
multi-parton interaction effects (MI),\cite{MI} whose presence has
been postulated at lower jet energies where the low-\xg\ cross
sections are often higher than expected.\cite{ZEUSJETS,H1MI} Studies
of the energy flow around the jets have shown that MI effects appear
to be small with the harder jets of the present study, giving 
stronger confidence that discrepancies between
the data and theory are not due to this cause.  Indeed in fig.\ 3 it is seen
that the agreement between experiment and theory is not perfect in all
kinematic regions.  Discrepancies are particularly evident with higher
incoming photon energies and when both emerging jets are at forward
$\eta$ values.  The existing models are unable to account for the
data, and more theoretical input would appear invited.

\section{Prompt photon measurements.}

\begin{figure}[t]
\centerline{
\parbox{5cm}{\hspace*{2mm}
\epsfig{file=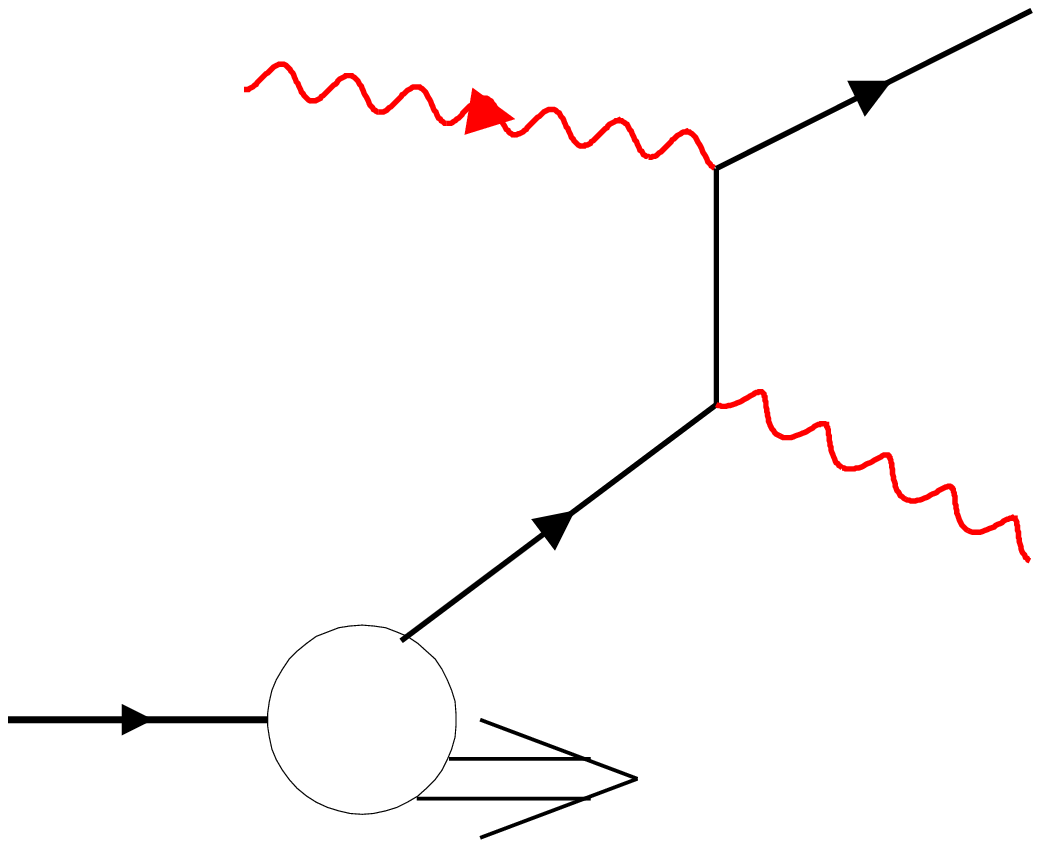,width=2.7cm,%
bbllx=80pt,bblly=90pt,bburx=400pt,bbury=440pt,clip=yes}
\\ \epsfig{file=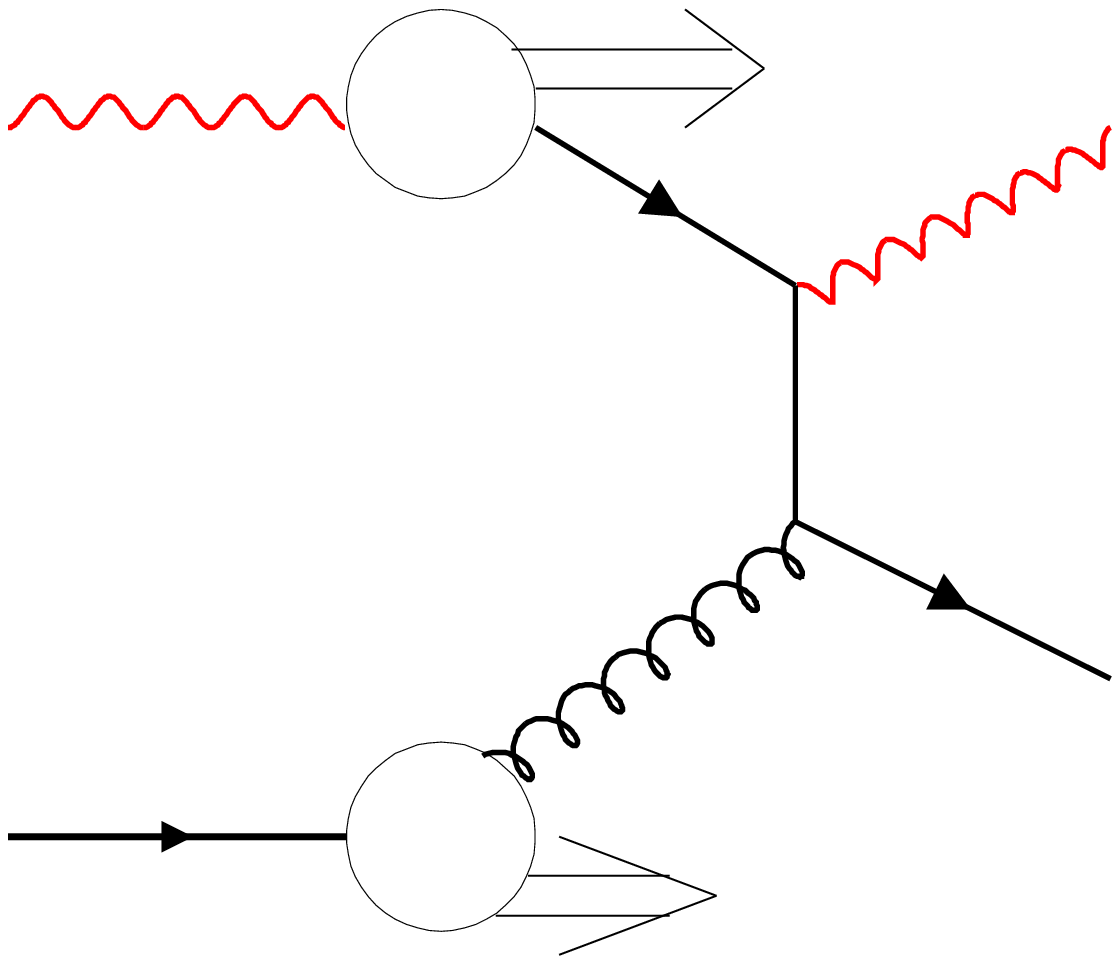,width=3.1cm,%
bbllx=55pt,bblly=100pt,bburx=400pt,bbury=440pt,clip=yes}
}\hspace*{1mm}
\raisebox{-3.3cm}{\epsfig{file=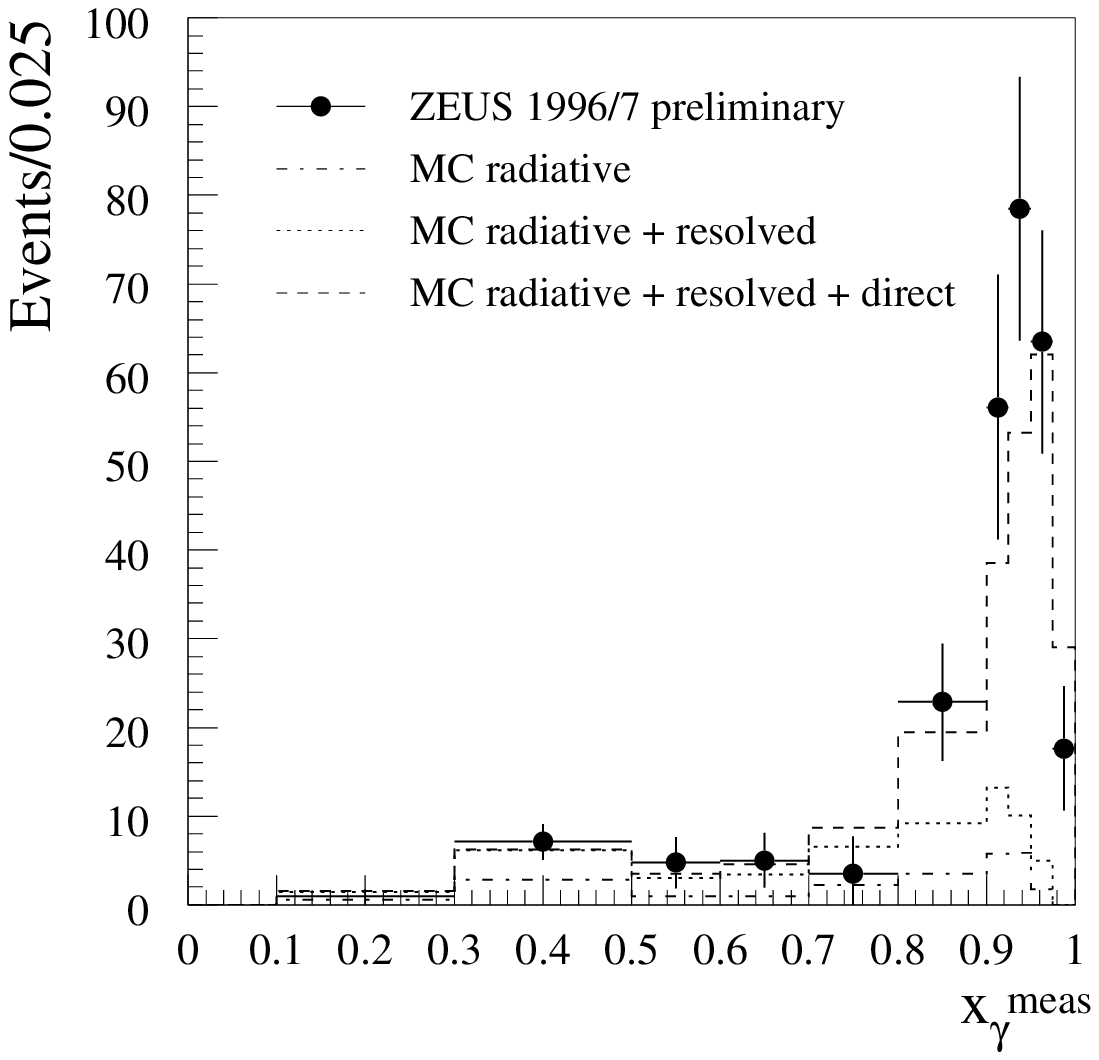,width=6.3cm,%
bbllx=0pt,bblly=0pt,bburx=320pt,bbury=320pt,clip=yes}
}}
\vspace*{-6mm}\hspace*{2cm}(a)\hspace*{6.1cm}(b)
\caption{(a) Examples of direct and resolved LO diagrams in prompt
photon photoproduction. (b) Distribution of $x_\gamma^{\;meas}$, 
defined at the detector level analogously to $x_\gamma^{\;jets}$, 
in photoproduced events having an isolated prompt photon and a
jet.  The photon satisfies $5 < E_T < 10 $ GeV and $-0.7 <
\eta < 0.9$; the jet satisfies $E_T > 4.5$ GeV
and is required within a central $\eta$ range.  The GRV~\protect\cite{GRV} 
photon structure is used.  
 }\end{figure}

As an alternative to jets, one can seek to measure the production of
high-$E_T$ photons, known as ``prompt
photons''.  Since QCD processes of this type have a different set of
diagrams from those of dijet processes (fig.\ 4(a)), they provide a
complementary perspective on the physics.  The photon is
already a partonic object; so there are fewer complications associated
with hadronisation (though a recoil jet is still present).  There is
less dependence on jet finding, and a photon is typically better
measured than a jet.  On the other hand, an important background from
hard neutral mesons (mainly $\pi^0, \eta$) must be subtracted, and
allowance must be made for photons radiated from high-$E_T$ quarks.
It is helpful to insist that the photon be isolated.  ZEUS impose an 
isolation condition such that within a cone of unit radius
in ($\eta, \phi$) around a photon of transverse energy $E_T$, no more
than a further $0.1E_T$ of transverse energy may be present.

\begin{figure}
\centerline{
\epsfig{file=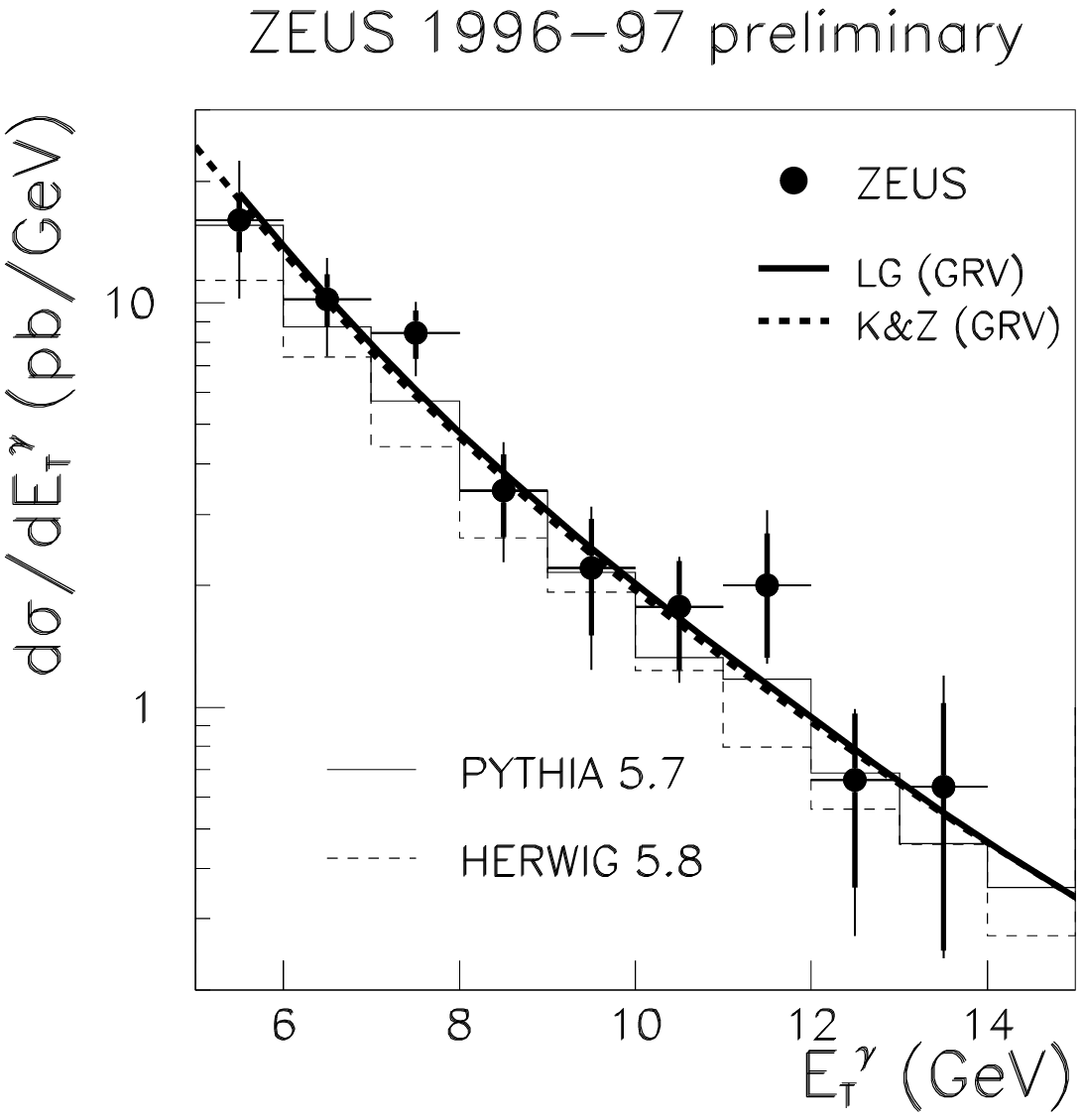,width=5.5cm%
,bbllx=0pt,bblly=0pt,bburx=320pt,bbury=340pt,clip=yes}\hspace*{7mm}
\epsfig{file=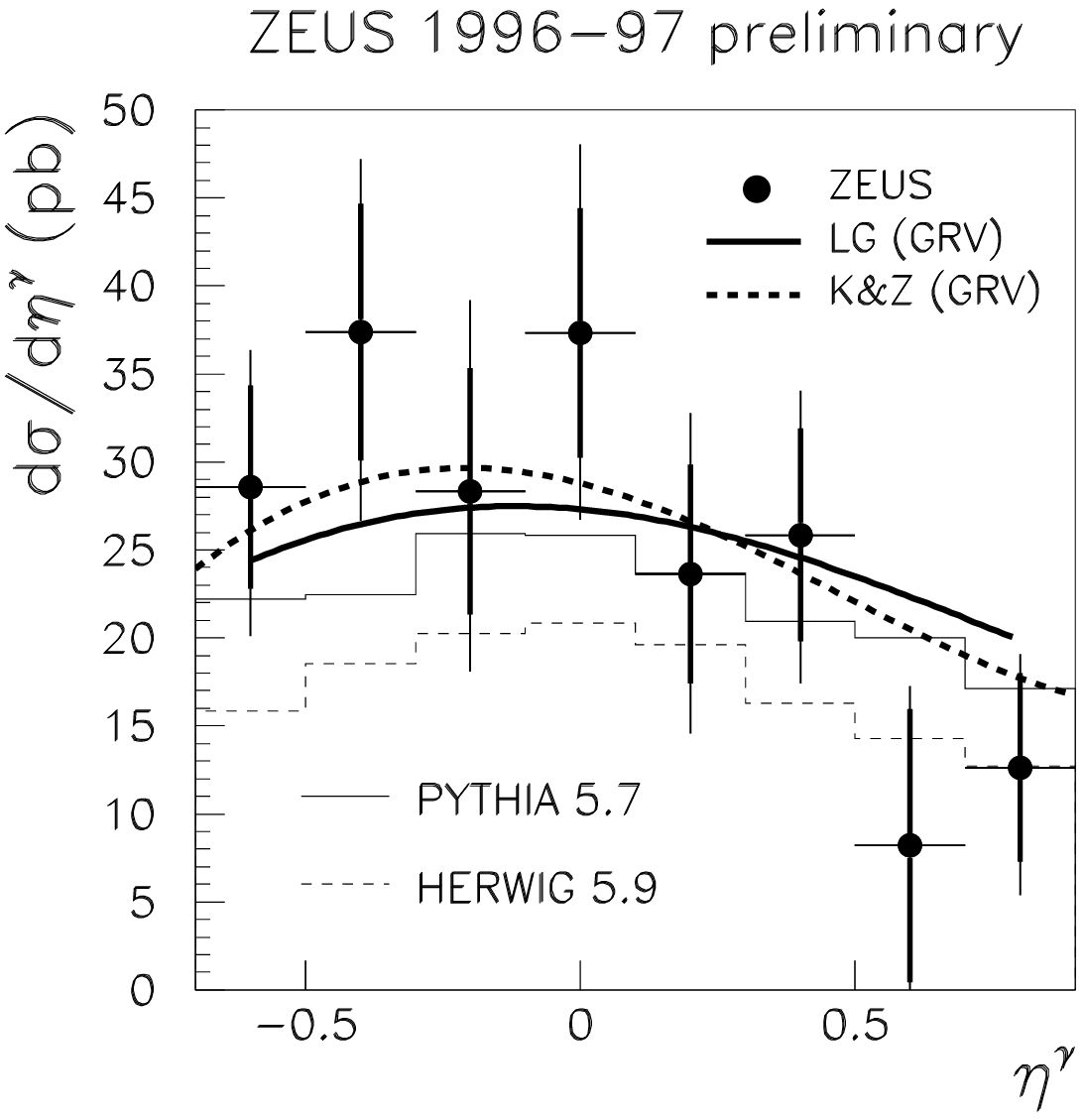,width=5.5cm%
,bbllx=0pt,bblly=0pt,bburx=320pt,bbury=340pt,clip=yes}
}
\vspace*{-5mm}\hspace*{2.5cm}(a)\hspace*{6cm}(b)
\caption{(a) $E_T$ distribution of inclusive prompt photons with $-0.7
< \eta < 0.9$. (b) $\eta$ distribution of inclusive prompt photons with
$5 < E_T < 10$ GeV.  Predictions are shown for PYTHIA
5.7, HERWIG 5.9 and two NLO calculations.\protect\cite{KZ,LG}}
\end{figure}

When a jet is also observed, a value of \xg\ may be measured
from the jet and the photon.  Figure
4(b) shows that the distribution is very different from that
found in dijet photoproduction.  The peak near unity, associated with
the direct diagrams, is considerably more pronounced than with dijets.
The distribution is quite well described using PYTHIA.  

Following an earlier paper on the observation of photoproduced prompt
photon processes,\cite{PRP} ZEUS have presented further preliminary
results.  The inclusive cross section as a function of $E_T$ is also
well described by PYTHIA (fig.\ 5(a)) and by NLO calculations,
although HERWIG appears low.  Given the large errors at present, the
available models describe the rapidity distributions satisfactorily
(fig.\ 5(b)).

\section{Gluon content of the photon.}
The hadronic coupling of the photon has to involve
charged components in the intermediate hadronic state, namely 
quarks -- but quarks should be 
accompanied by a gluon component.  In most models, the 
quarks tend to take more of the photon momentum than the gluons, 
so that a search for gluons in the photon should aim at a sensitivity 
at low \xg\ values. 

\begin{figure}
\centerline{\epsfig{file=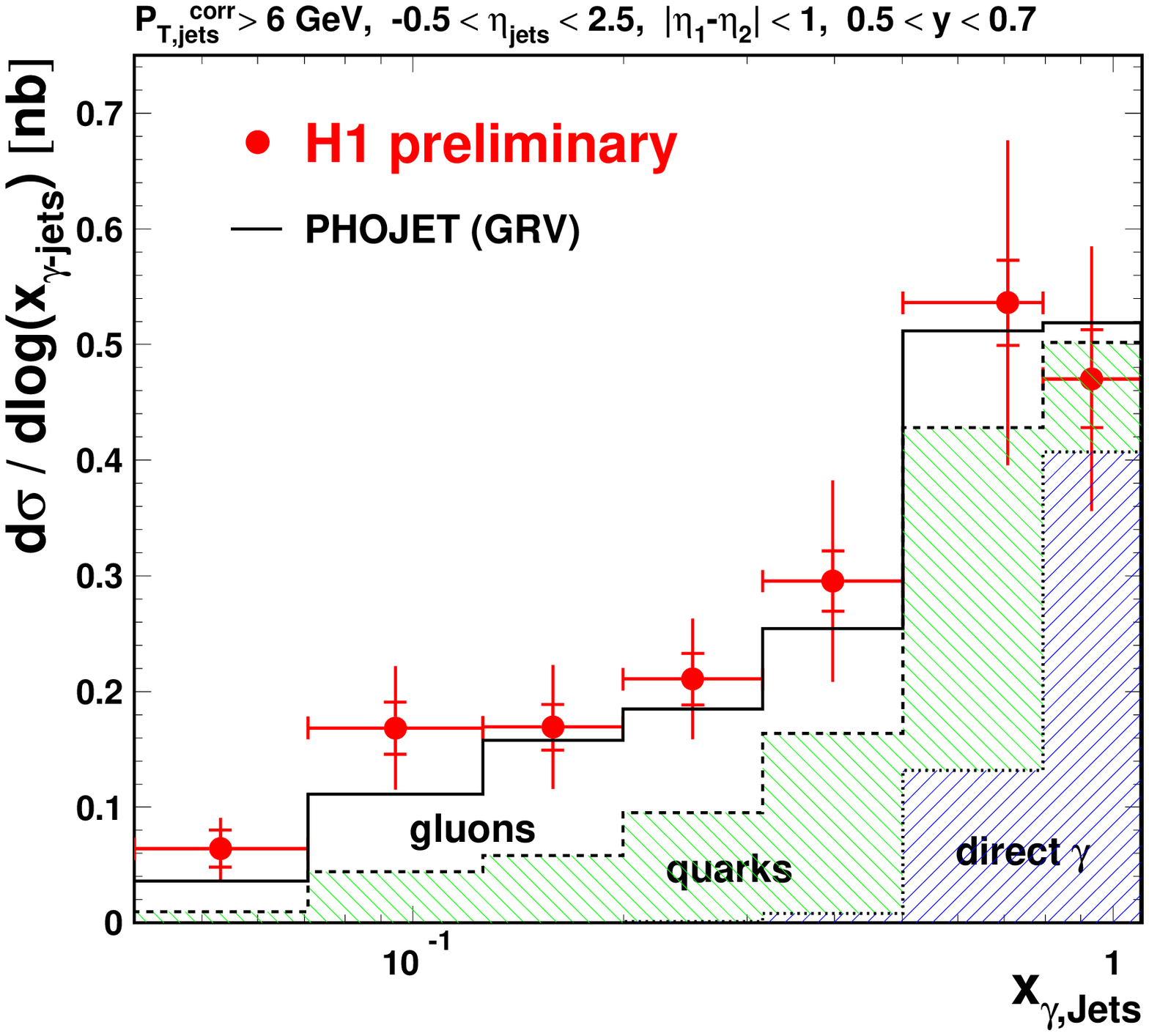,width=5.4cm}\hspace*{5mm}
\epsfig{file=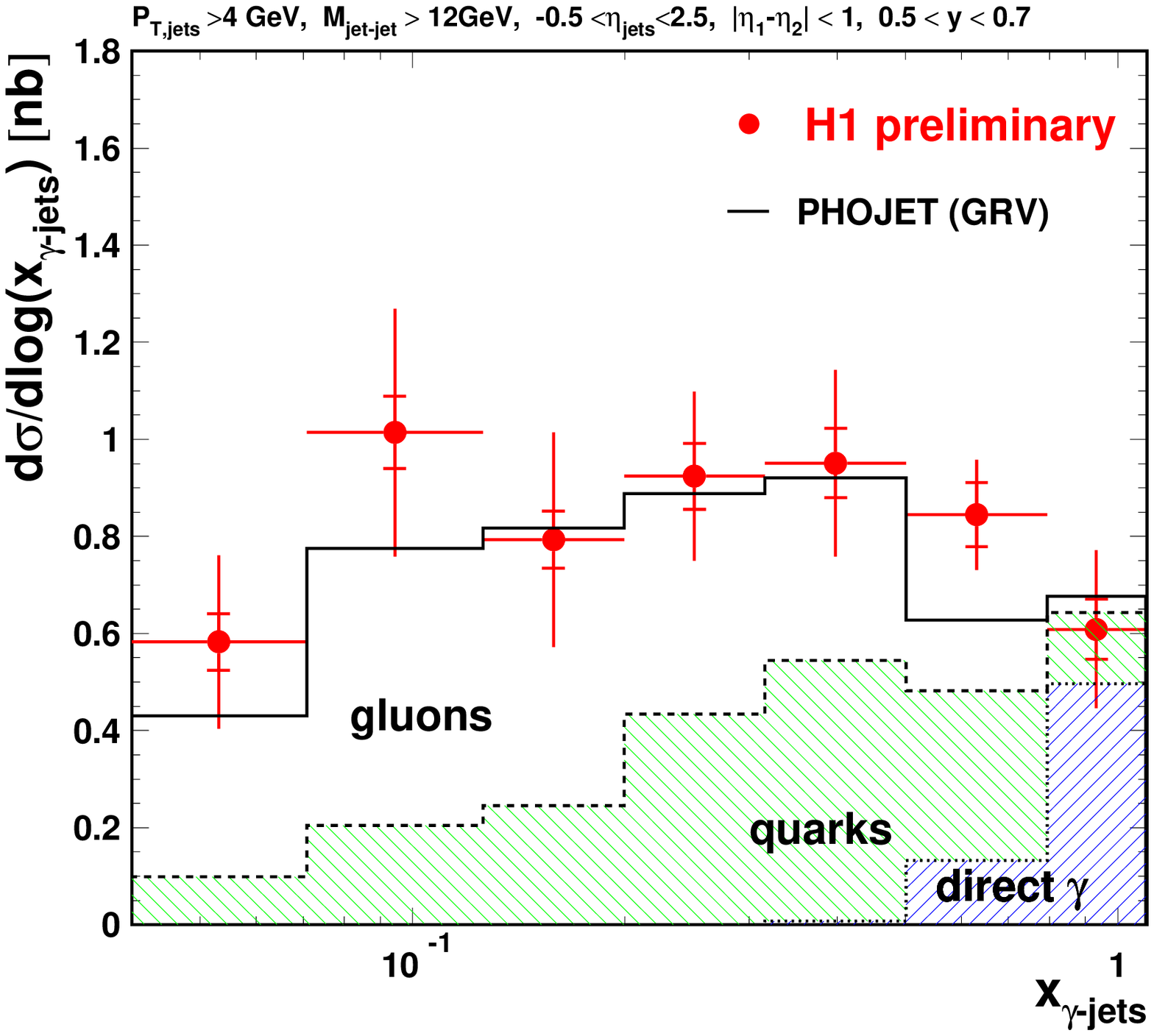,width=5.4cm}
}
\vspace*{-4mm}\hspace*{2.8cm}(a)\hspace*{5.8cm}(b)
\caption{Plots of $x_\gamma$ (H1) using dijets of mass greater than 12 GeV.  
In the (a), the minimum jet $E_T$ is 6 GeV, in (b) it is 4 GeV.}\end{figure}

\begin{figure}[b!]
\centerline{
\epsfig{file=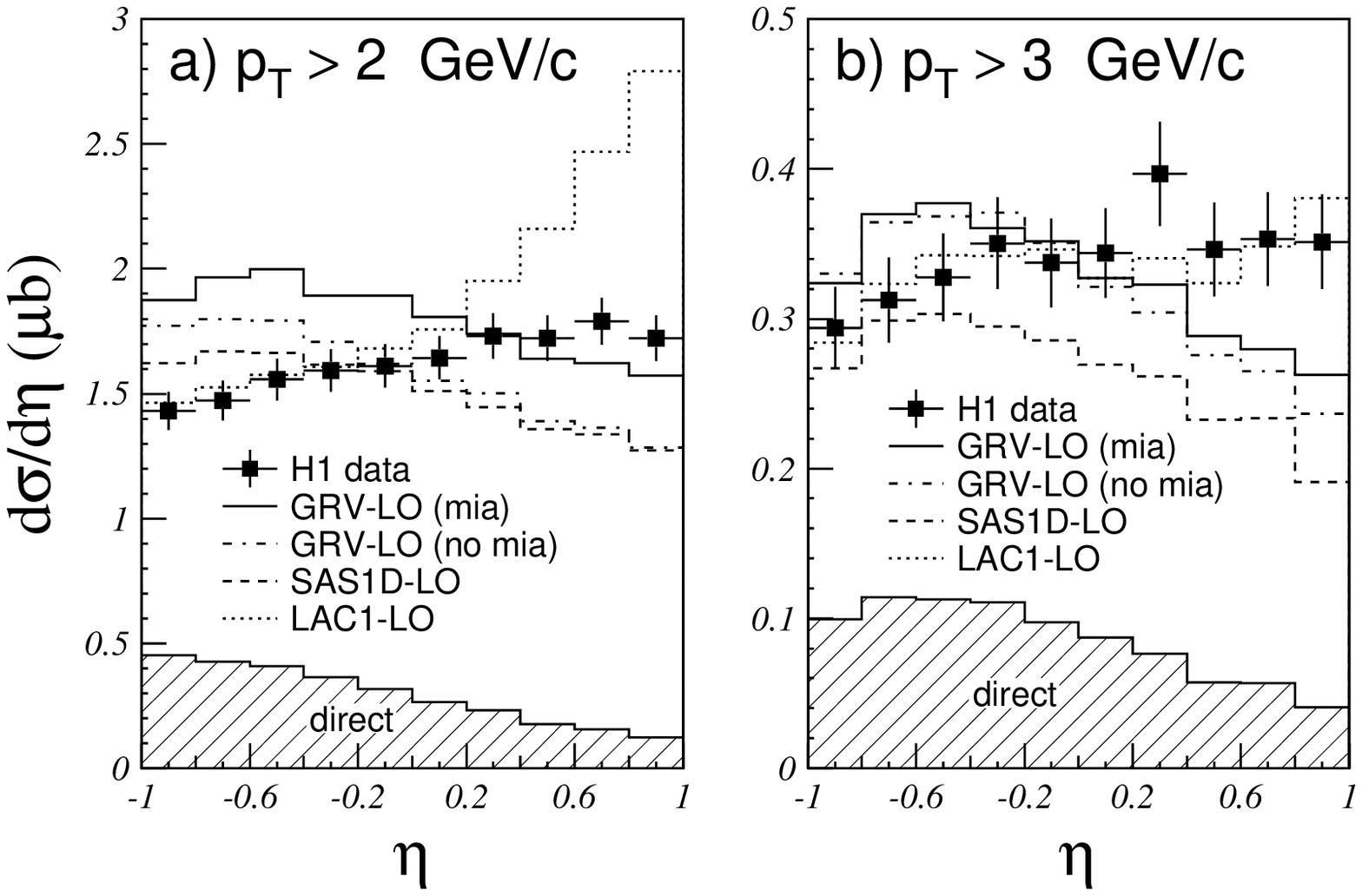,width=6.3cm,%
bbllx=50pt,bblly=265pt,bburx=550pt,bbury=580pt,clip=yes}\hspace*{-6.5mm}
\epsfig{file=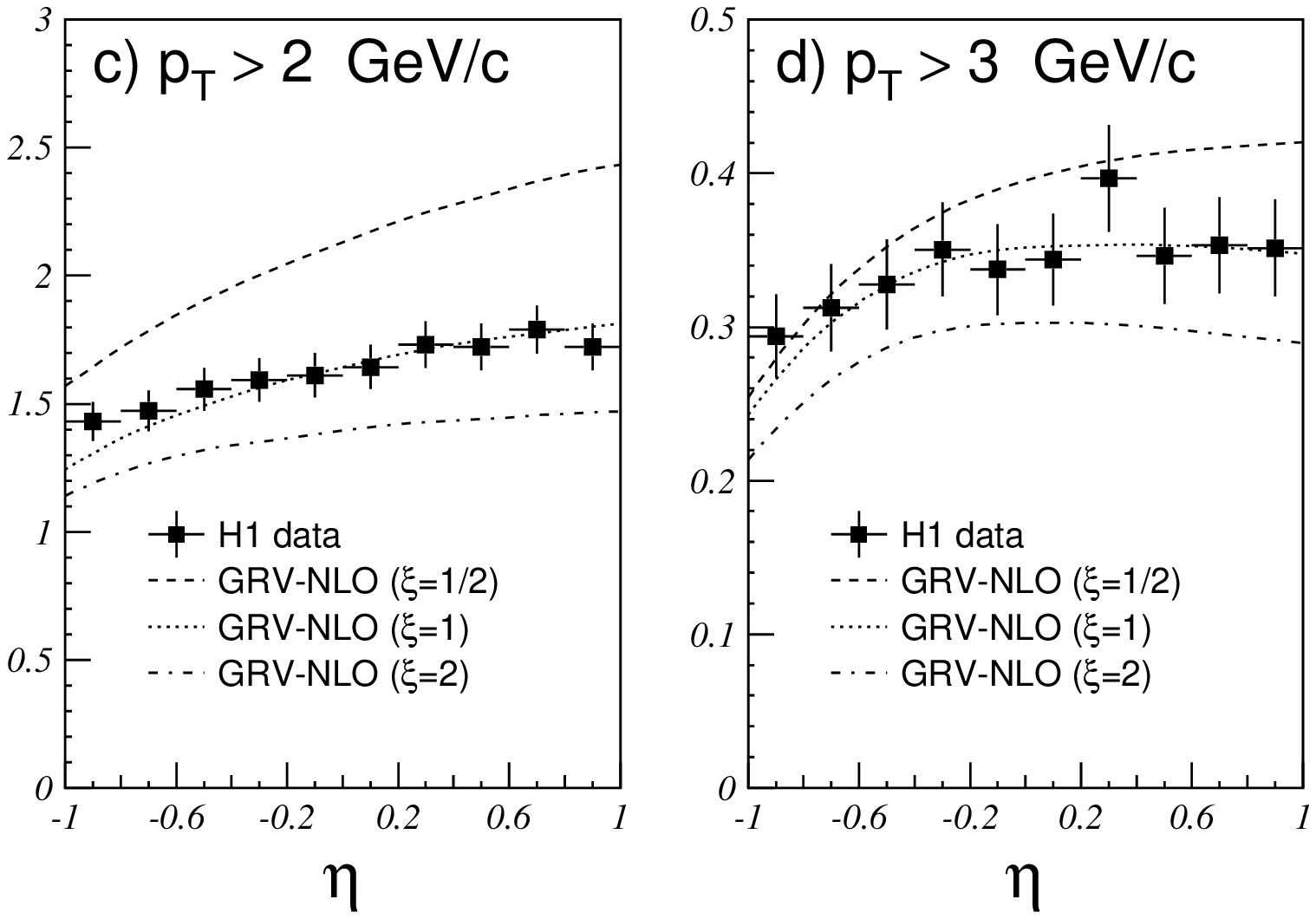,width=6.3cm,%
bbllx=50pt,bblly=265pt,bburx=550pt,bbury=580pt,clip=yes}}
\caption{Distribution of inclusive charged particles measure by H1.
The data are compared with (a,b) LO Monte Carlo distributions and
(c,d) NLO curves.}
\end{figure}

This is illustrated in fig.\ 6, which shows \xg\ distributions from
dijet events in H1.  The shape of the distribution is changed
radically when jets with lower $E_T$ values are accepted, even while
the minimum dijet mass remains at the same value.  The Monte Carlo
based predictions comprise components from direct processes, and from
resolved processes initiated by quarks and by gluons within the
photon.  This partition of the resolved component is illustrated using
the GRV photon parton densities,\cite{GRV} which have a generally good
record and here too give a good fit to the data.  Note that although
the direct peak is no longer visible as such in (b), this is entirely
a consequence of  the much higher rates now obtained at lower \xg\
values. Through accepting the lower-$E_T$ jets, one has greater acceptance
at low \xg\ giving a greater sensitivity to the gluon.  Indeed, if
the quark part of the resolved photon is as modelled above, the level
and shape of the cross sections already imply a quantitative
observation of gluons in the photon.

A further study has now been performed by means of inclusive charged
particles measured in H1.\cite{H1gluon} Above transverse momenta of
2--3 GeV, such particles are fairly well associated with high-$p_T$
partons; they are also well measured in the apparatus.  Distributions
are shown in fig.\ 7, where it is demonstrated that the 3 GeV data can
be described by means of LO theory using a suitably chosen photon
structure.  NLO calculations are also successful, but there exists a
large QCD scale uncertainty.

An \xg\ estimator $x_\gamma^{rec} = \sum p_Te^{-\eta}/E_\gamma$ is now
evaluated by summing over selected high-$p_T$ tracks in an event.
Using PYTHIA, good correlation is found between $x_\gamma^{rec}$ and
\xg\ at the LO parton level, which permits an unfolding procedure 
to be used to evaluate cross sections as a function of \xg.  These are
then converted to photon parton densities. Calculated quark
densities in the photon are now subtracted off (with a small amount of
model dependence), and one is left with a measurement of the gluon
densities.

The assumption behind this procedure is that the existing photon
models, based on $e^+e^-$ collider data, have relatively
well-determined quark densities.  The H1 results are shown in fig.\
8, together with those from a similar recent analysis which uses jets
rather than charged particles.  The two sets of points are consistent,
and agree well with the GRV model of the photon gluon
density.  As is commonly the case, LAC1~\cite{LAC} is too high.

\begin{figure}\centerline{
\epsfig{file=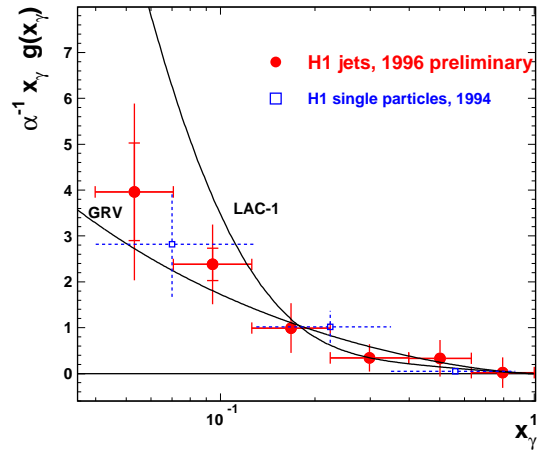,width=7cm}}
\caption{Gluon density in photon, as measured by H1 using 
inclusive tracks and dijet events. The error bars include
uncertainties in the subtraction of the quark
component.}\end{figure}

\newpage
\section{Studies of the virtual photon}
The LO differential cross section for the process $ep\to $ dijets
can be written
\begin{eqnarray} \lefteqn{\hspace*{-10mm}\frac{d^5\sigma}
{dy\,dx_\gamma\,dx_p\,d\cos\theta^*\,dQ^2}= 
\frac{1}{32\pi s_{ep}\, y x_\gamma x_p} \times} \nonumber \\ 
& & \moreht\times\sum\,f^k_{\gamma/e}(y,Q^2)
\,f^k_{i/\gamma}(x_\gamma,P_t^{2},Q^2) \,f_{j/p}(x_p,P_t^{2})
|M_{i,j}(\cos\theta^*)|^2.
\end{eqnarray}
The sum is over the polarisation $k$ of the photon emitted by the
electron, and the parton species $i$ and $j$ in the photon and the 
proton, respectively.  The terms $f$ denote the photon density in the 
electron and the parton densities in the photon and proton; $M_{i,j}$
denotes the QCD matrix element.

Within the acceptance of a typical experiment, most of the $2\to2$
parton processes are $t-$channel exchanges with similar shape.  The
Single Effective Subprocess approximation~\cite{SES} replaces all the $M_{i,j}$
terms by a common expression $M_{SES}(\cos\theta^*)$, and forms
combinations $\tilde{f}$ of the parton densities.  These are known as
Effective Parton Densities (EPDs).  Taking into account the colour
charge on the gluon, one obtains for the proton:
$$
 \tilde{f}_p = \sum
\left(f_{q/p}(x_p,P_t^2) + f_{\bar q/p}(x_p,P_t^2)\right) + 
\textstyle\frac{9}{4}f_{g/p}(x_p,P_t^2), 
$$
summing over quark flavours.  Only transverse polarised photons are
retained.  Then the sum in (1)can be replaced by
the single effective term:
$$
\hspace*{\fill}
 \,f^T_{\gamma/e}(y,Q^2)
\,\tilde{f}_{\gamma}(x_\gamma,P_t^{2},Q^2) \,\tilde{f}_{p}(x_p,P_t^{2})
|M_{SES}(\cos\theta^*)|^2.
$$

\begin{figure}
\centerline{\epsfig{file=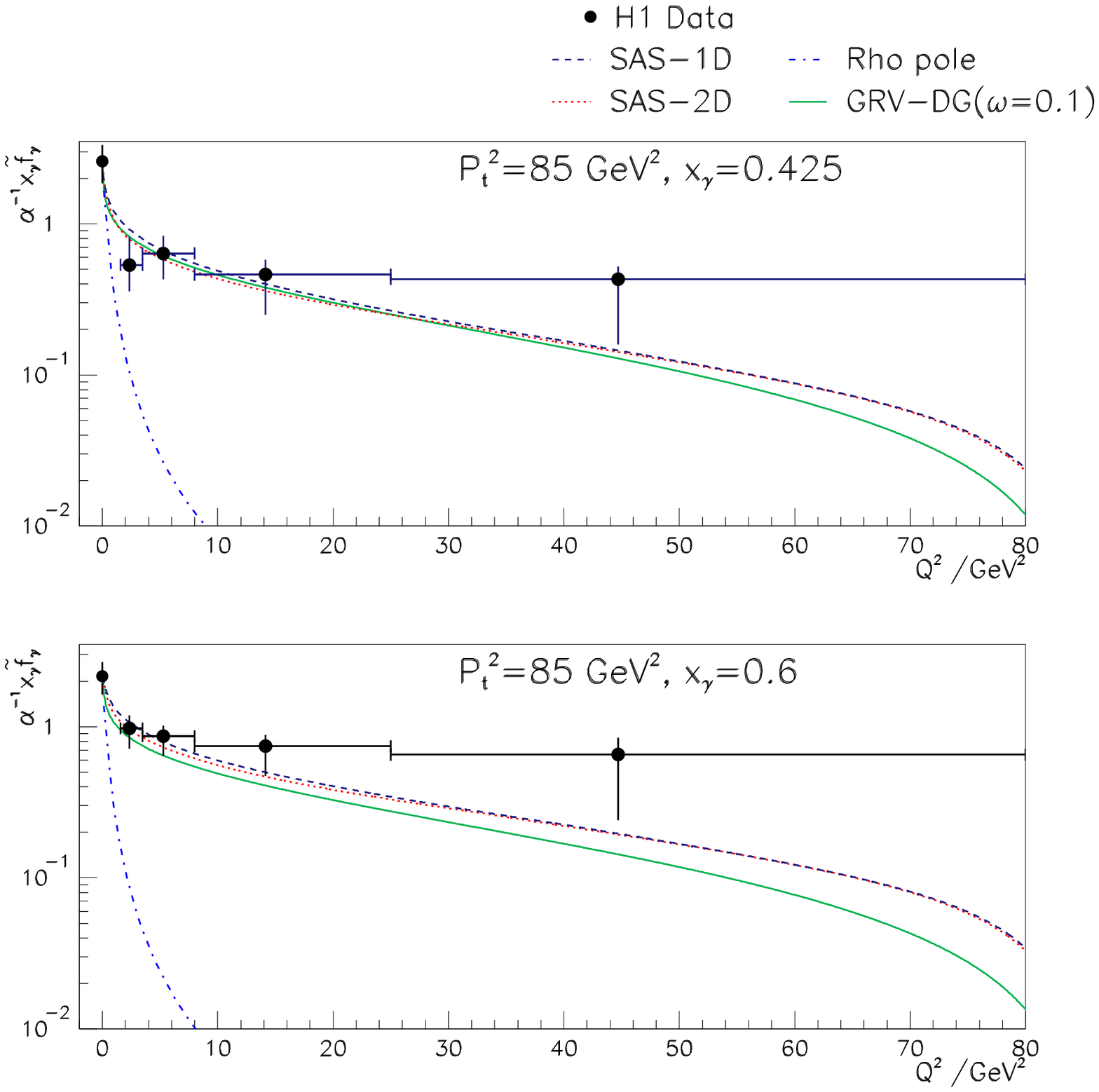,width=6.7cm%
,bbllx=20pt,bblly=30pt,bburx=545pt,bbury=547pt,clip=yes}
\hspace*{6mm}\raisebox{-1mm}{\epsfig{file=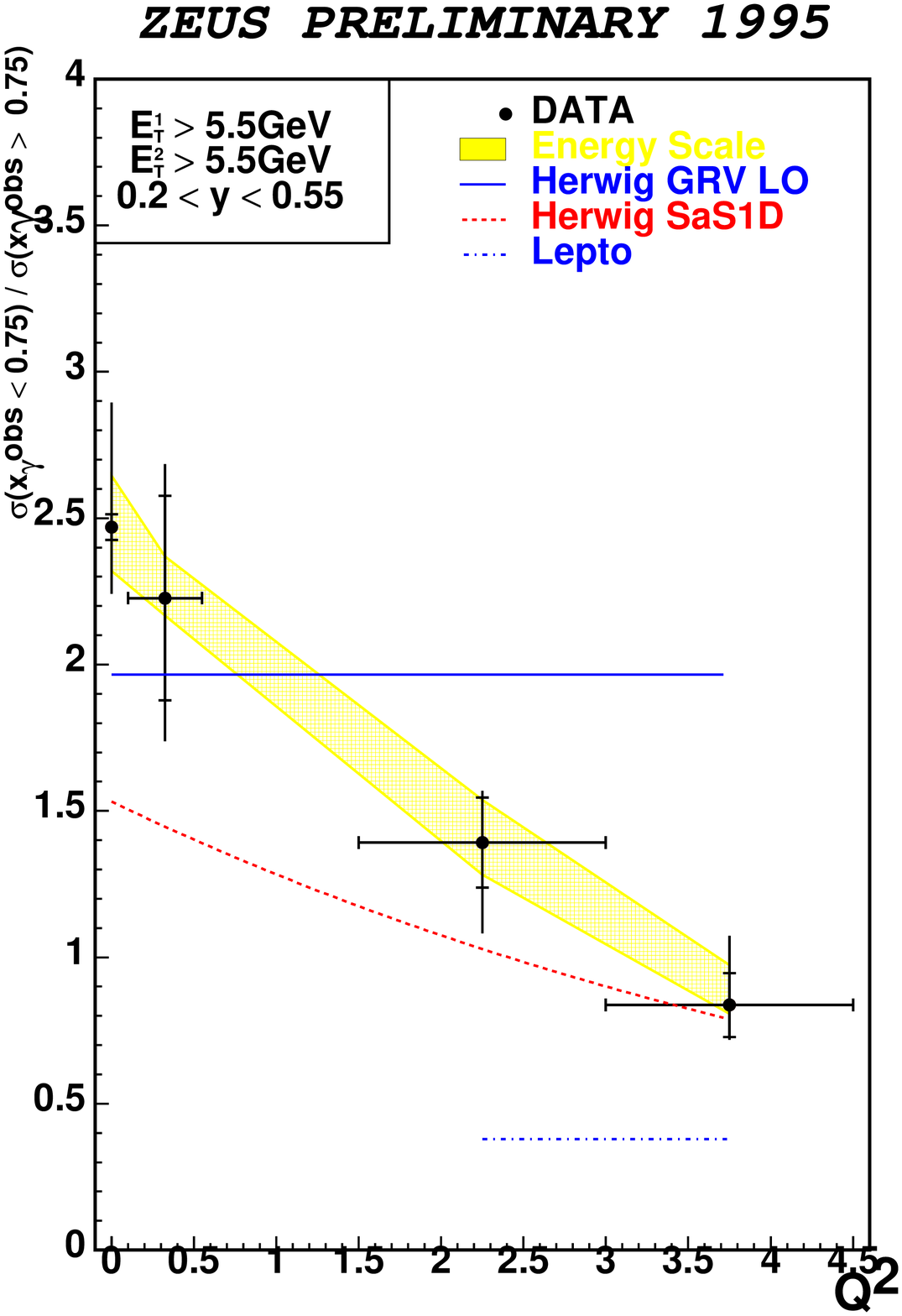,width=4.7cm
,bbllx=-10pt,bblly=0pt,bburx=583pt,bbury=640pt,clip=yes}
}}
\vspace*{-2mm}\hspace*{2.9cm}(a)\hspace*{6cm}(b)
\caption{$Q^2$ dependence of (a) effective parton densities in photon (H1)
and (b) ratio of low-$x_\gamma^{obs}$ to high-$x_\gamma^{obs}$
cross sections (ZEUS).}
\end{figure}
The SES and EPD approximations provide an experimental procedure for
comparing the properties of the photon in different situations.  
On this basis, H1 have measured the EPD
$\tilde{f}_\gamma$ in the photon over a range of small to medium
values of $Q^2$, using photoproduced jet pairs of transverse
momentum $P_t$.\cite{H1virt}  The aim is to examine the variation of
$\tilde{f}_\gamma$ with $x_\gamma$, $Q^2$ and $P_t^2$ and compare with
QCD predictions. 

This is a two-scale situation.  It is first found that the
$x_\gamma^{jets}$ distributions show a clear trend, as $Q^2$ and
$P_t^2$ separately increase, to be more and more dominated by the
``direct peak'' at $x_\gamma^{jets}\approx 1$.  In other words, the
hadronic properties of the photon decrease as the process becomes
``harder''.  Such a situation is not new and has commonly involved the
employment of meson-like form factors.  However the HO QCD diagrams
add further complexity and interest to the present type of
interaction.

Figure 9(a) shows the variation of $\tilde{f}_\gamma$ with $Q^2$ at two
\xg\ values.  (The variation with $P_t^2$ is found to be
weak.)  Also plotted are predictions from a pure VMD $\rho$-pole form
factor, and modern QCD-based calculations from Schuler and
Sj\"ostrand.\cite{SaS} The latter are valid only in the kinematic
range $\Lambda_{QCD}^{\;2} \ll Q^2 \ll P_t^2$.  Two conclusions
follow; one is that as now expected, a $\rho$ pole form factor alone 
is quite insufficient.  The second is that the QCD approach gives good
results in the region where its validity is claimed, namely for
$0.1 \ll Q^2 \ll 85 \mbox{GeV}^2$ but perhaps fails 
elsewhere.  Altogether, given the approximations made, this would seem
to represent a considerable success, and confirms the H1 collaborations earlier results in terms of inclusive jets.\cite{H1JETQ2}

Over a finer scale of low $Q^2$ values, however, ZEUS have shown that
at present there still may be problems in describing the behaviour of
the photon.  Fig.\ 9(b) shows the cross section ratio
$\sigma(x_\gamma^{obs} < 0.75)/\sigma( x_\gamma^{obs} > 0.75)$ for a
kinematic range of centrally produced jet pairs.  At $Q^2 \approx 0$
the lower $x_\gamma^{obs}$ range is dominated by hadronic
(``resolved'') diagrams.  The ratio falls with $Q^2$, as predicted by
SaS, but the latter theory does not reproduce the magnitude of the
effect.  At $Q^2\approx 0$, the data lie above the GRV prediction
(which is not modelled to fall with $Q^2$), indicating the possible
presence of multiparton interactions under these conditions.  The
Lepto line represents a model with no hadronic component to the
photon, and NLO effects confined to parton showers; it fails completely 
at these $Q^2$ values.

\section{Study of the photon remnant.}
\begin{figure}
\centerline{\epsfig{file=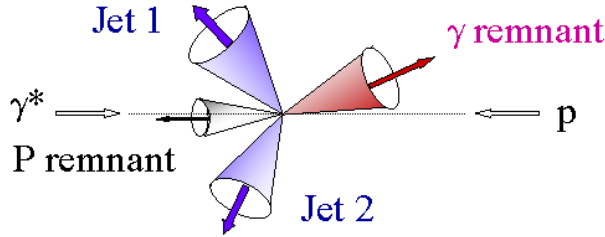,width=9cm}}
\caption{Event topology for photon remnant studies.}
\end{figure}
The properties of the photon remnant offer a means for testing whether
the so-called anomalous photon processes are correctly described in
present theories.  To achieve this, H1 have employed a customised jet
finder to identify events with two high-$E_T$ jets accompanied by a
proton remnant and a photon remnant.  The latter is then treated as a
jet-like object whose properties can be studied.  Fig.\ 11 shows the
mean $p_T$ of the photon remnant relative to the beam, plotted as a
function of (a) the virtuality of the incident photon, and (b) the mean
\ET\ of the two hard jets.  In this way, theory can be
tested over a range of kinematic parameters.

The results show that HERWIG describes the behaviour of the photon
remnant quite well, apart perhaps from the case of the
softest ``hard'' jets.  An intrinsic $k_T$ of partons within the
photon of 0.66 GeV was taken.  Within HERWIG, most
of the transverse momentum observed at $Q^2\approx 0$ arises from the
effects of hadronisation: to study the parton level properties of the
anomalous photon coupling, it may therefore be best to go to $Q^2$ values of more
than a few GeV$^2$.  On the other hand, the RAPGAP Monte Carlo is not
successful in this context.



\begin{figure}\centerline{
\epsfig{file=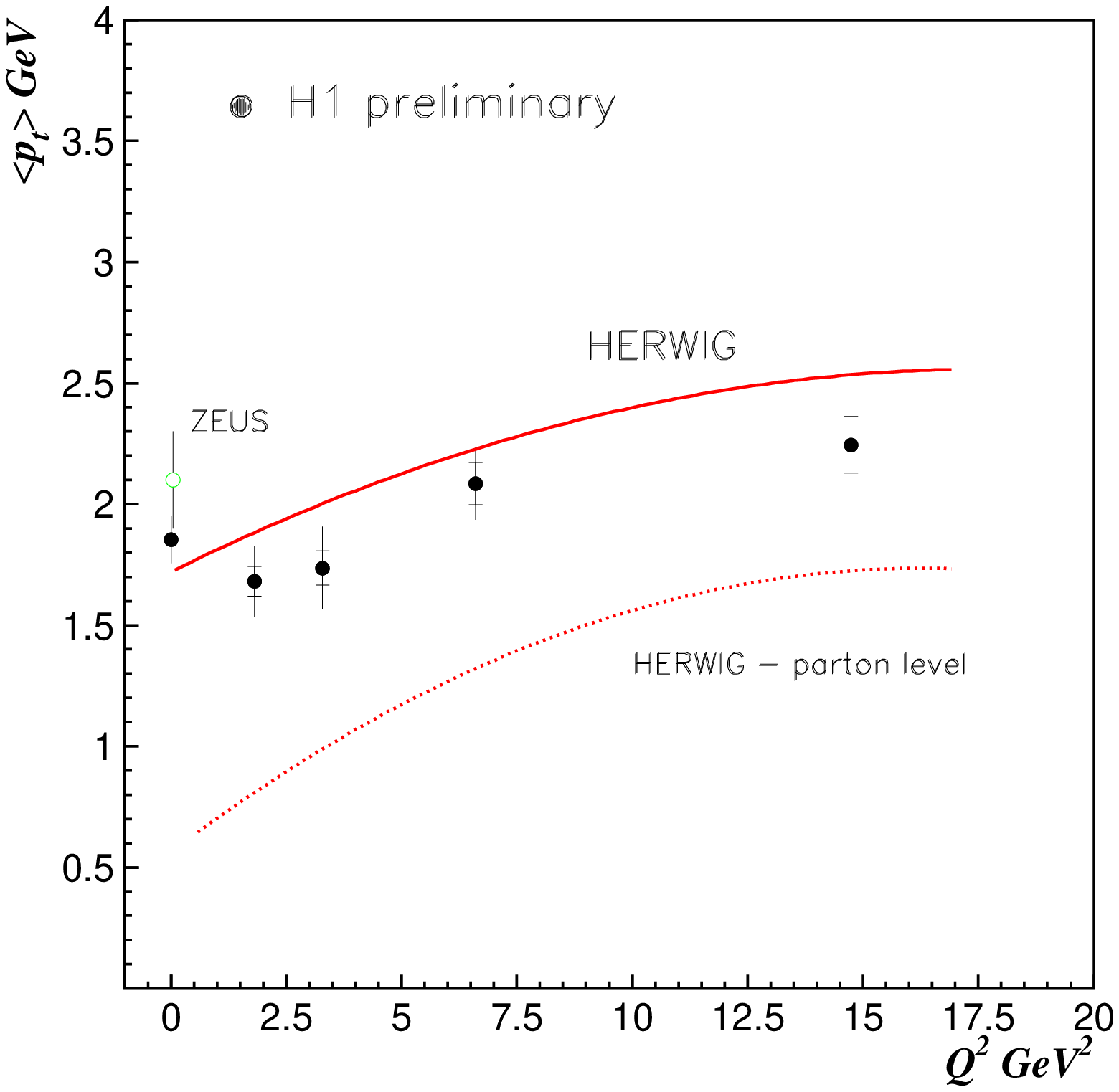,width=5.8cm%
,bbllx=40pt,bblly=190pt,bburx=500pt,bbury=630pt,clip=yes}
\raisebox{-2mm}{\epsfig{file=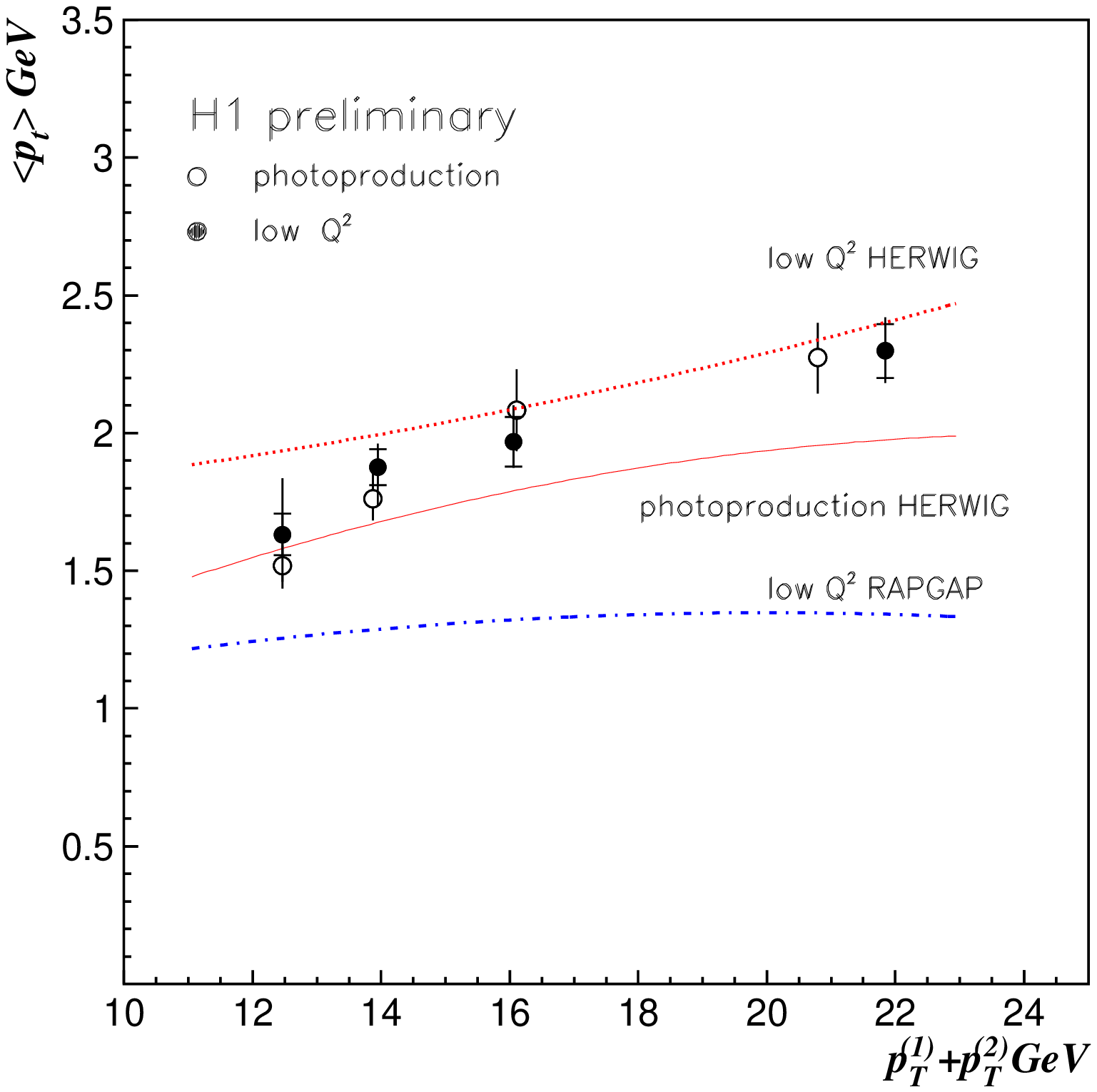,width=6.2cm%
,bbllx=30pt,bblly=180pt,bburx=510pt,bbury=630pt,clip=yes}}\hspace*{2mm}}
\vspace*{-3mm}\hspace*{3.1cm}(a)\hspace*{5.9cm}(b)
\caption{Variation of the mean $p_T$ of the photon remnant
(a) with the virtuality of the photon (b) with the mean transverse
energy of the hard jets.}
\end{figure}

\section{Three-jet events in photoproduction.}
With the aim of confirming our understanding of the QCD processes that
are operative in the photoproduction of jets, ZEUS have performed an
investigation of three-jet events in photoproduction.\cite{Esther}  The most
important experimental requirement was for the three-jet mass to
exceed 50 GeV.  The kinematics of the final state are best studied in
the three-jet centre-of-mass frame, as illustrated in fig.\ 12.  The
angle $\psi_3$ is defined between the plane of the three jets, and the
plane containing the incident proton and photon directions and the
highest-$E_T$ jet, labelled ``3''.

The distribution of $\psi_3$ is sensitive to a number of kinematic and
dynamical properties of the process (fig.\ 12).  It should first be
noted that owing to the kinematic requirements on the jets, the
distribution is depleted near zero and $180^\circ$; what
appears as a two-peaked structure would otherwise be a steep-sided
valley.  Likewise, the flat $\psi_3$ distribution that would be had
if the three jets (or high-$p_T$ partons) were produced according to
phase space now becomes centrally peaked.  The latter shape in no way
resembles the data, confirming that dynamical mechanisms are shaping
the three-jet production.  

Higher-order calculations describe the shape of the observed
distribution well, as indeed do HERWIG and PYTHIA.  From further
investigations it was found that the main contribution to the
three-jet process comes from initial-state gluon radiation from the
proton and photon. (The higher peak near $180^\circ$ comes from the
higher amount of initial-state radiation from the proton.)
Final-state radiation is relatively suppressed since the third jet
then needs to become substantially separated from its parent jet.  It
is also possible to switch off the colour coherence structures within
PYTHIA.  The result is a flatter $\psi_3$ distribution whose shape
does not agree well with the data.  With further statistics, one may
hope to distinguish between the PYTHIA and HERWIG models and possibly
some of the higher order calculations.

\begin{figure}
\centerline{
\epsfig{file=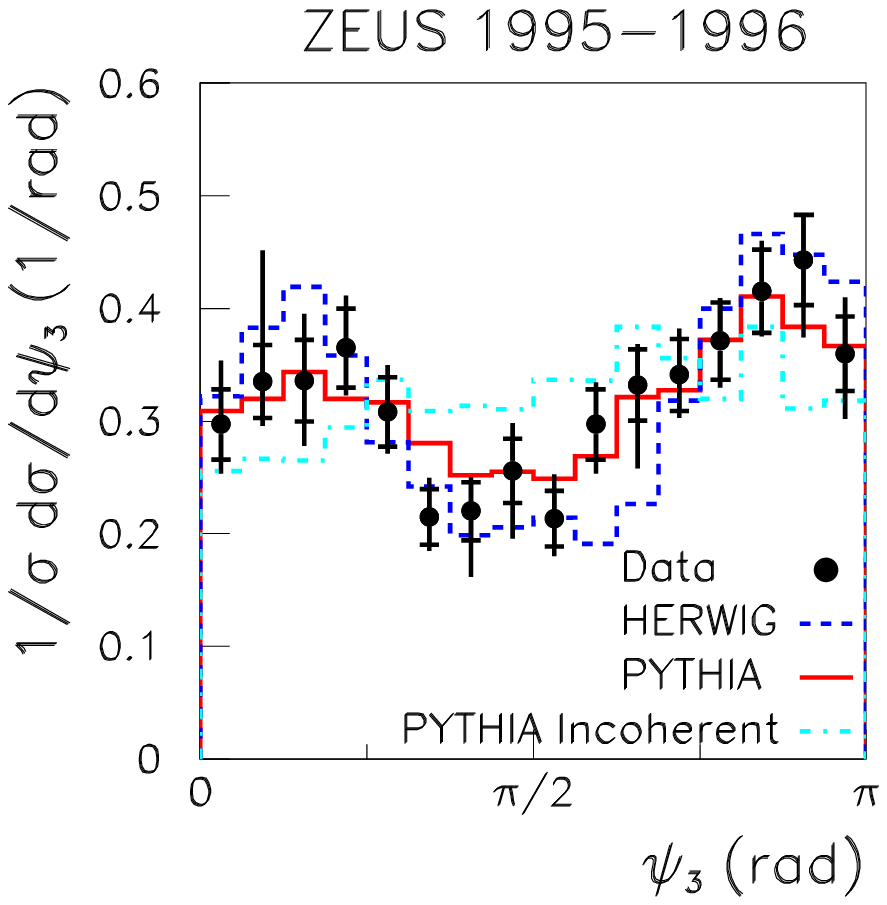,width=5.4cm}
\epsfig{file=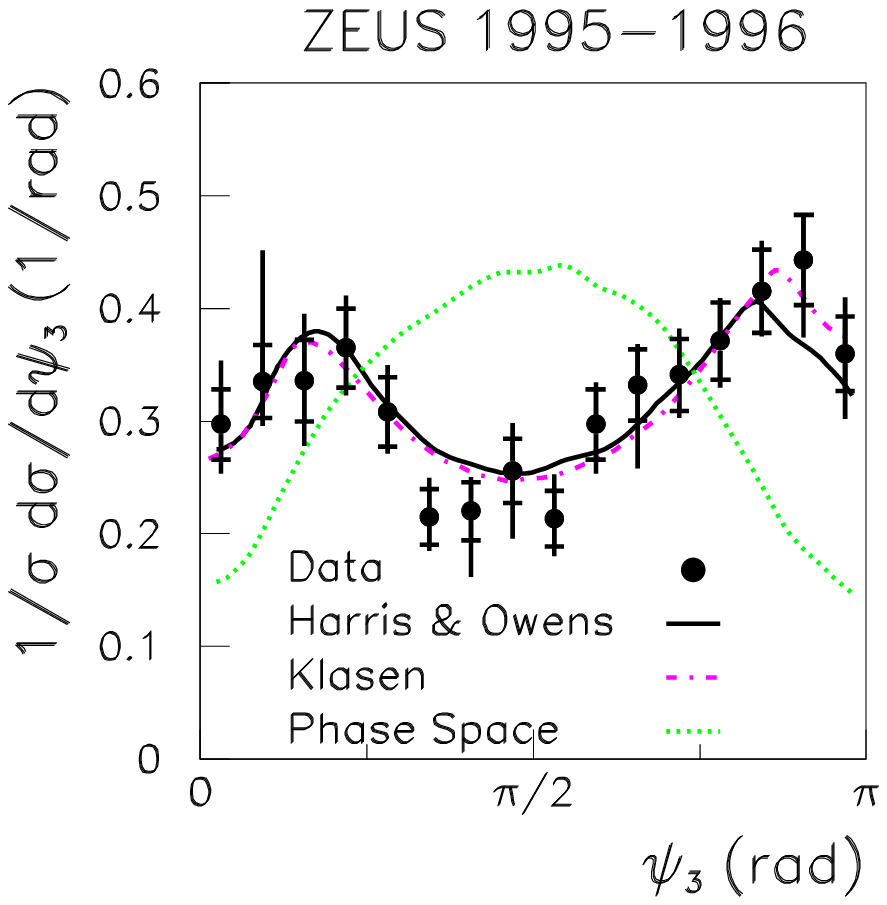,width=5.4cm}}
\centerline{\epsfig{file=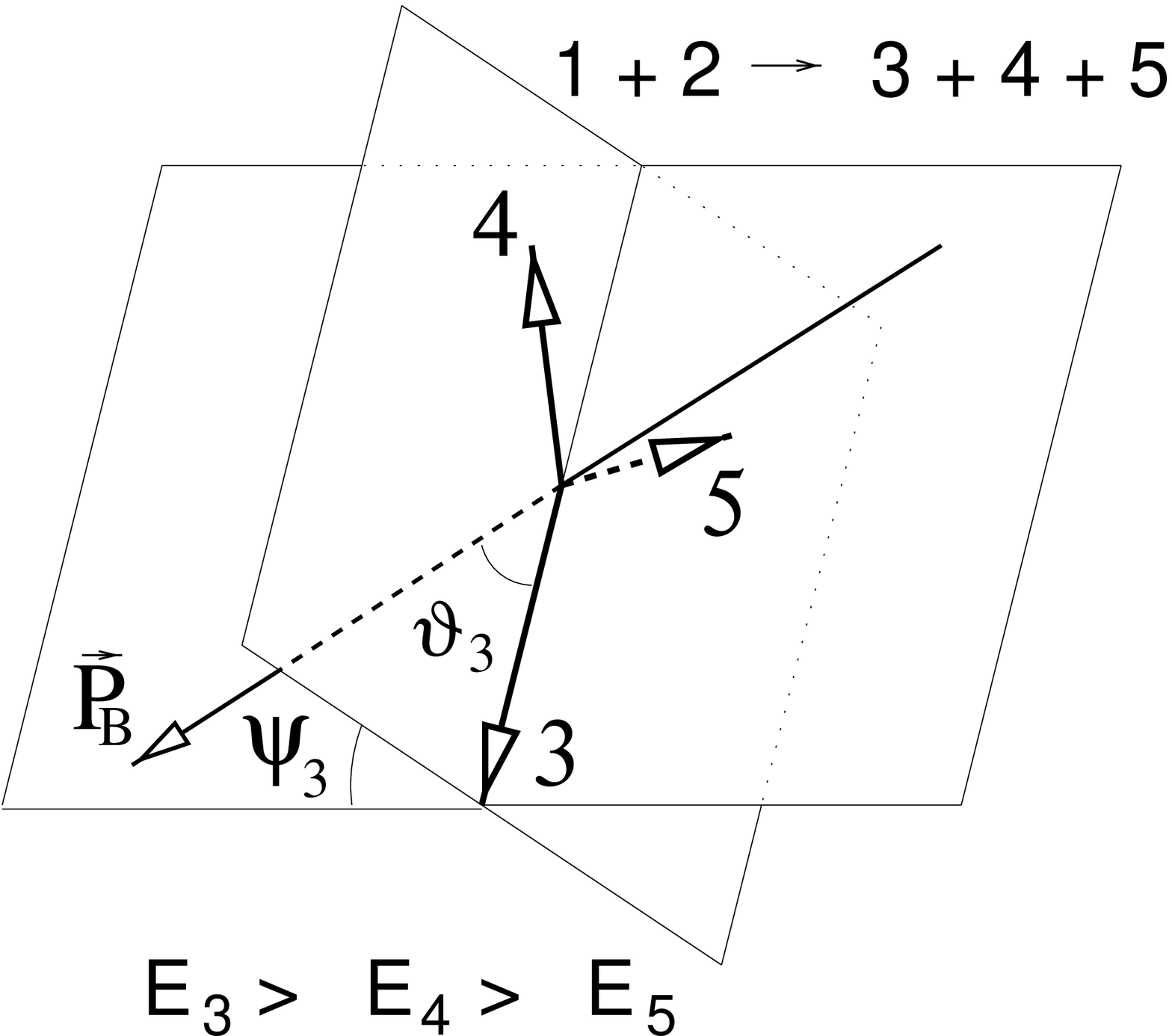,width=5cm}}
\caption{Definition and distribution of $\psi_3$ in three-jet 
events in photoproduction.}
\end{figure}

\section{SubJets}
A further useful tool for studying the parton structure of the photon
would be a knowledge of the nature of a given hard jet.  When a jet is
due to a heavy quark, this may often be determined on a jet-by-jet
basis; however the majority of jets in photoproduction are from light
quarks and gluons.  Studies made by ZEUS of jet widths~\cite{JETW}
have now been supplemented by studies on the numbers of ``subjets''
within a jet.  The concept of a subjet arises in terms of clustering
jet algorithms such as the $k_T$ algorithm, where a cut $y_{cut}$ on
the clustering parameter determines whether two jet candidates shall
be merged. The algorithm is iterated until all remaining pairs of jet
candidates have clustering parameters above this value, which is
chosen to correspond to an acceptable overall jet radius.  A large
value of $y_{cut}$ gives more merging, and fewer final jets.

\begin{figure}
\centerline{
\epsfig{file=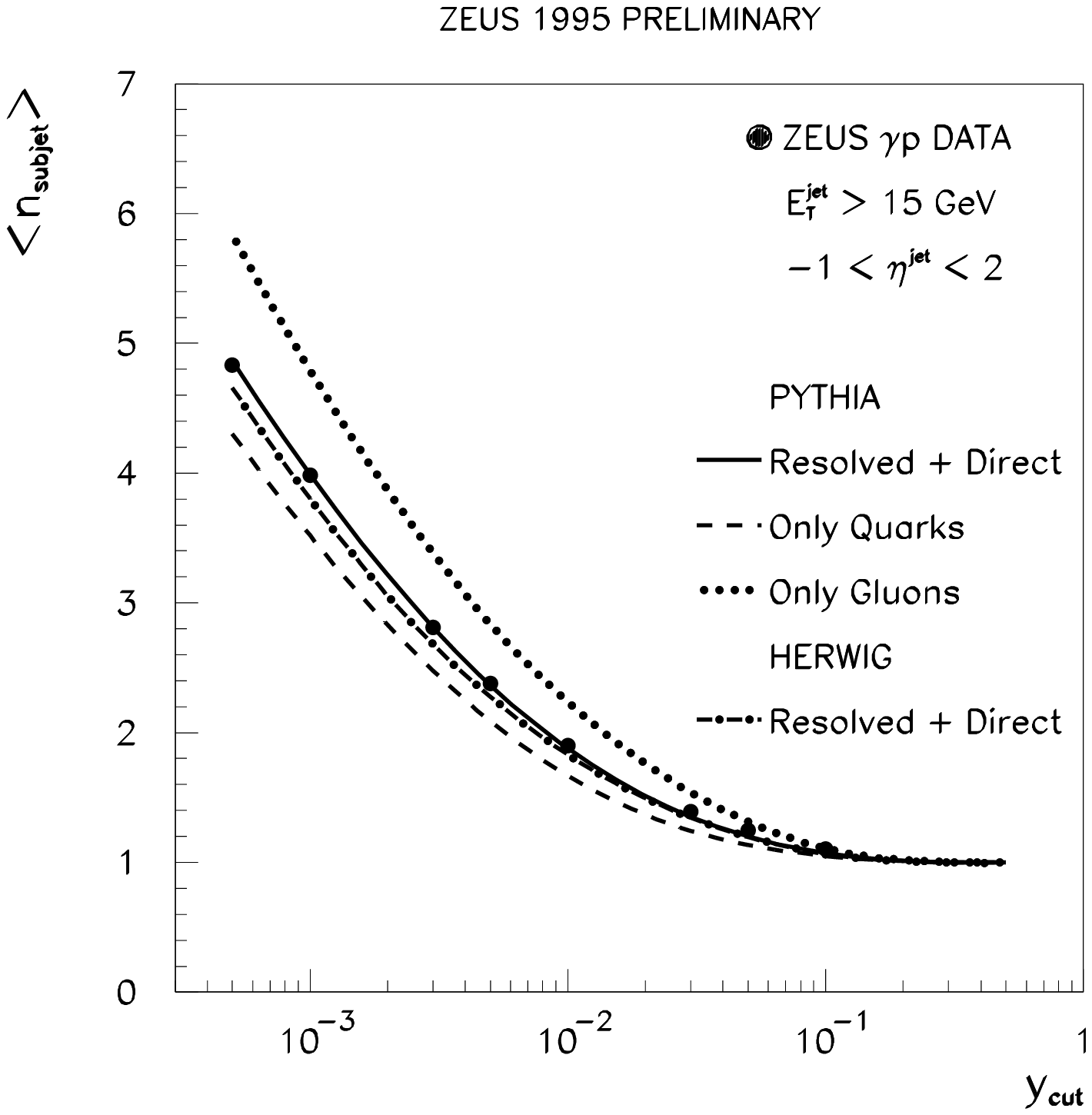,width=5.6cm%
,bbllx=50pt,bblly=70pt,bburx=455pt,bbury=485pt,clip=yes}
\hspace*{4mm}\epsfig{file=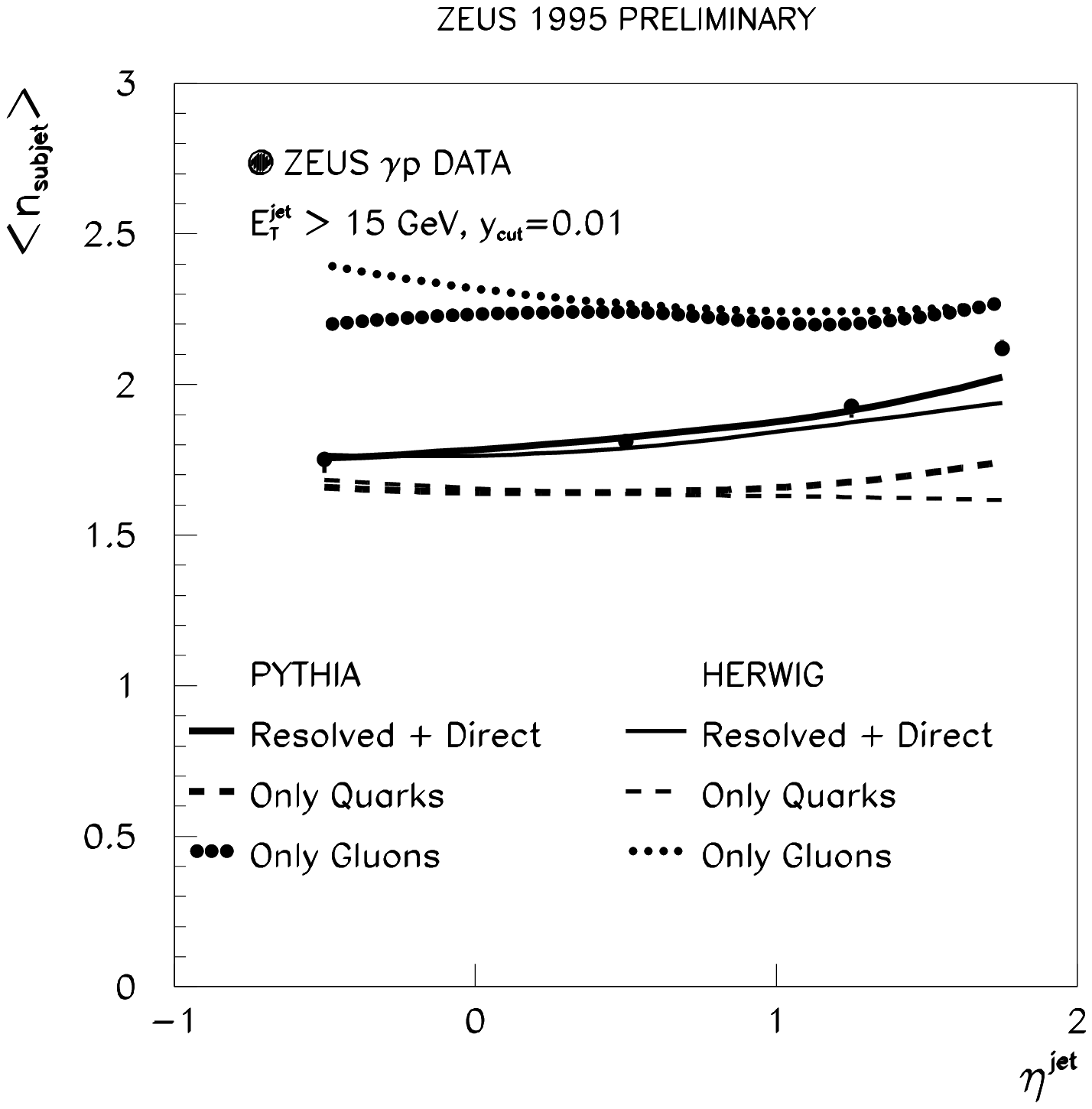,width=5.6cm%
,bbllx=50pt,bblly=70pt,bburx=455pt,bbury=485pt,clip=yes}\hspace*{2mm}}
\vspace*{-3mm}\hspace*{2.9cm}(a)\hspace*{6cm}(b)
\caption{(a) Variation of number of subjets with cut parameter 
for photoproduced jets with $E_T > 15$ GeV at HERA (b) variation with
pseudorapidity for fixed $y_{cut} = 0.01$.
Gluon jets are predicted to have higher \nsubj\ than quark jets.}
\end{figure}

Having found a jet, it is possible to rerun the algorithm using
varying smaller values of $y_{cut}$; a number of smaller jets
(``subjets'') may now be obtained in place of the original one.  This
number, at given $y_{cut}$, is an indicator of the internal structure
of the jet.

Fig.\ 13(a) shows the variation of the mean number of subjets
\nsubj\ with $y_{cut}$ for photoproduced high-$E_T$ jets
in ZEUS. The form of the variation can be simulated using
HERWIG and PYTHIA, and is intermediate between the expectations for
pure quark-initiated and pure gluon-initiated jets, corresponding well
to an appropriate mixture of direct and resolved processes.  

With a chosen fixed $y_{cut}$ value, it is then possible to study the
variation of \nsubj\ with, for example, the pseudorapidity of the
primary jet, as shown in fig.\ 13(b).  It is apparent that the jets
are predominantly quarklike at negative $\eta$ values but
predominantly gluonlike for positive $\eta$.  The variation of the
data is reasonably well reproduced using the GRV photon structure.  It
is too early to say if this technique can prove fruitful in
distinguishing between different photon models, but it is clearly a
potentially useful addition to the physicist's repertoire in an area
where investigations are notoriously difficult.

\section{Conclusions}
HERA provides an environment in which many aspects of photon physics
can be investigated. The past two years have seen a significant expansion
both in the numbers of photoproduction analyses and in their scope.
Studies of jet production have reached the stage where kinematic
regions can be investigated where the possible complications from
underlying parton-parton events should not be a problem; nevertheless,
there are discrepancies even here with the present theoretical
predictions, even at next-to-leading order, which seem to invite new
thinking at the theoretical level.  The study of prompt photons in
photoproduction is now a possibility, given the present possible
luminosities at HERA, and is able to provide new perspectives in the
area of hard photon interactions.

The gluon content of the photon has been studied further by the
extension of existing techniques.  These results from H1 are
consistent with the best available photon models, but the errors
remain fairly large.  Both H1 and ZEUS have begun to study the
transition from the quasi-real photon, with its extensive hadronic
properties, to the virtual state, still at moderate $Q^2$ values,
where ``anomalous'' photon coupling becomes a more dominant effect,
providing an perturbative QCD element in low-$x_\gamma$ photon
physics.  New studies of the photon remnant provide further insight
into the ``anomalous'' photon.

The actual QCD mechanisms remain an important area of study.  Here,
multi-jet final states and the teazing apart of jets into ``subjets''
have begun to give improved confirmation of our understanding of
detailed aspects of the hadronic interactions initiated by a photon.

To conclude, photon physics at HERA has entered a new and remarkably
productive stage.  The entire area of heavy flavour production was
omitted here, being part of another speaker's remit!  There are still
many avenues to investigate and many questions to answer. The future
at HERA will hopefully provide us with the opportunity to unravel
further the properties of a particle, the photon, which proves to have a
richer and more complex nature the more it is examined.

\section*{Acknowledgments}
The author wishes to thank members of the H1 and ZEUS collaborations
for much helpful information in the preparation of this talk, and also
the Royal Society (London) for financial assistance in attending the
Conference.

 \end{document}